\documentclass[prb,twocolumn,amsmath,amssymb, letterpaper]{revtex4-1}

\usepackage{amsmath}
\usepackage{amssymb}
\usepackage{amsfonts}
\usepackage{graphicx}
\usepackage{dcolumn}
\usepackage{float}
\usepackage{bm}
\usepackage{mathrsfs}
\usepackage{txfonts}
\usepackage{tabularx}
\usepackage{bigstrut}
\usepackage[usenames,dvipsnames]{color}

\begin{document}

\title{ Coherent control of optical injection of spin and currents in topological insulators }

\author{Rodrigo A. Muniz and J. E. Sipe }

\affiliation{Department of Physics and Institute for Optical Sciences, University
of Toronto, Toronto ON, M5S 1A7, Canada}

\begin{abstract}
Topological insulators have surface states with a remarkable helical
spin structure, with promising prospects for applications in spintronics.
Strategies for generating spin polarized currents, such as the use
of magnetic contacts and photoinjection, have been the focus of extensive
research. While several optical methods for injecting currents have
been explored, they have all focused on one-photon absorption.

Here we consider the use of both a fundamental optical field and its
second harmonic, which allows the injection of spin polarized carriers
and current by a nonlinear process involving quantum interference between one- and two-photon
absorption. General expressions are derived for the injection
rates in a generic two-band system, including those for one- and two-photon
absorption processes as well as their interference. Results are given
for carrier, spin density and current injection rates on the surface
of topological insulators, for both linearly and circularly polarized
light. We identify the conditions that would be necessary for experimentally
verifying these predictions. 
\end{abstract}

\date{{\small{{{\today}}}}}

\maketitle

\section{Introduction }

Three-dimensional topological insulators are fascinating materials,
with a band gap in the bulk and protected midgap states on their surfaces
\cite{hasan10,qi11}. The surface electronic bands are described by
a single Dirac cone with a helical spin structure, which is the equivalent
of a dominant Rashba spin-orbit coupling term in the Hamiltonian.
This property leads to a number of interesting features, including
non-magnetic scattering, the magnetoelectric effect \cite{qi08,essin09},
and the formation of Majorana fermions in the proximity of superconductors
\cite{fu08}. Due to the effective spin-orbit coupling, the spin and
current of the surface states are closely related \cite{raghu10},
providing an exciting opportunity for technological applications using
spin polarized currents. There have already been several studies using
the proximity of a magnetic metal for injecting spin polarization
and current \cite{butch12,modak12,semenov12,mahfouzi12}.

Another fruitful approach for manipulating currents in materials involves
optical excitation. The optical properties of topological insulator
surface states are very interesting themselves, with features such
as the injected current depending explicitly on the Berry phase \cite{tse10,hosur11}.
The injection of spin and current by one photon absorption processes
has been studied in different circumstances \cite{hosur11,misawa11,mciver12nn}.
In order to break the rotational symmetry stemming from the Dirac
cone - a necessary step for generating a current - the use of an in-plane
magnetic field, the application of strain, and an oblique angle of
incidence have all been considered. Corrections due to snowflake warping
have been included; even a surprisingly relevant contribution from
the Zeeman coupling of the light field has been identified \cite{refael13}.
Nonlinear effects due to the second harmonic have also been considered
\cite{hsieh11a,hsieh11b,mciver12prb,sobota12}, especially in the
treatment of pulses. However the focus of even these studies has been
on one photon absorption processes.

One of the most interesting techniques for optical injection is coherent
control, an example of which involves tuning the interference of
one and two photon absorption processes to achieve a target response.
This has been employed for injecting carriers, spin polarization,
currents, and spin currents in semiconductors \cite{rioux12,kiran11},
and currents in graphene \cite{norris10,rioux11,kiran12}. It has
even been proposed that it could be used to inject a macroscopic Berry
curvature in semiconductor quantum wells \cite{virk11}. Here we present
predictions of the optical injection of carrier density, spin polarization,
charge current, and spin current at the surface of a topological insulator.
In order to identify the fundamental properties of coherent control
in topological insulators, we use a Hamiltonian with a perfectly symmetric
Dirac cone, and restrict the analysis to light at normal incidence.
This also helps to contrast the results of coherent control with those
obtained by other means. We keep a $\sigma_{z}$ mass term in the
Hamiltonian in order to analyze the dependence on the Berry phase,
which has interesting effects on the injection rates.

In Sec. \ref{sec:response} we present the calculation of optical
injection rates for an arbitrary quantity using Fermi's golden rule,
considering one and two photons absorption processes as well as their
interference. In Sec. \ref{sec:twoband} we provide general expressions
for the injection rates of a generic two-band system, especially for
carrier density, spin density, charge current and spin current operators.
Since two-band models can be used, as a first approximation, to compute
optical properties of a large number of materials, the expressions
derived there should be of use even beyond their application to topological
insulators. In Sec. \ref{sec:topins} we apply the results of Sec.
\ref{sec:twoband} to topological insulators. In Sec. \ref{sec:results}
we present the results for linearly and circularly polarized light,
referring to the Appendices \ref{app:formula} and \ref{app:coeffs}
for details. In Sec. \ref{sec:discussion} we end with a discussion
of interesting features in our results and the possibilities for their
experimental verification, including estimates for the expected experimental
results. Since the experimental techniques required to confirm our
results are well established, we can expect that such experiments
will help advance the understanding and applications of topological
insulators.

\section{Response to light fields }

\label{sec:response}

There are several methods for computing the response of a system to
external perturbations; one of the simplest and most standard methods
is Fermi's golden rule. It is especially suitable for coherent control
calculations because it makes evident all the contributions stemming
from one- and two-photon processes and their interference. This is
a feature not shared by the Kubo formalism, for instance.

The calculation for the injection rates of operators using Fermi's
golden rule has been already well explained in previous studies \cite{rioux12}.
However, it has been typically assumed that the fundamental photon
energy is below the bandgap, as is the case for most studies of semiconductors.
Since we will deal with systems that are gapless, there will be an
additional interference term. And in order to make the notation clear
we will present the main steps of the full calculation.

The full Hamiltonian in the presence of the external perturbation $V_{ext}\left(t\right)$
is $H\left(t\right)=H_{0}+V_{ext}\left(t\right)$, where $H_{0}$
is the Hamiltonian without the perturbation. The wavefunction in the
presence of the external perturbation can have a contribution from
an excitation of a valence $v$ band electron to a conduction $c$
band 
\begin{equation}
\begin{array}{rl}
\left|\Psi\left(t\right)\right\rangle = & c_{0}\left|0\right\rangle +c_{cv,\boldsymbol{k}}\left(t\right)\left|cv,\boldsymbol{k}\right\rangle +\dots,\\
\rho_{cv,\boldsymbol{k}}= & \left|c_{cv,\boldsymbol{k}}\right|^{2},
\end{array}
\end{equation}
where $\left|0\right\rangle $ is the groundstate of $H_{0}$ with
filled valence bands; $\left|cv,\boldsymbol{k}\right\rangle =a_{c,\boldsymbol{k}}^{\dagger}a_{v,\boldsymbol{k}}\left|0\right\rangle $
is the state with an electron-hole pair, with $a_{c,\boldsymbol{k}}^{\dagger}$
denoting the electron creation operator, and 
\begin{equation}
\begin{array}{rl}
c_{cv,\boldsymbol{k}}\left(t\right)= & \int_{-\infty}^{t}\frac{dt_{1}}{i\hbar}e^{-i\omega_{cv}t}\vartheta_{cv,\boldsymbol{k}}^{\left(1\right)}\left(t_{1}\right)\\
 & +\int_{-\infty}^{t}\frac{dt_{1}}{i\hbar}\int_{-\infty}^{t_{1}}\frac{dt_{2}}{i\hbar}e^{-i\omega_{cv}t}\vartheta_{cv,\boldsymbol{k}}^{\left(2\right)}\left(t_{1},t_{2}\right),
\end{array}\label{eq:coefC}
\end{equation}
where $\omega_{cv}=\omega_{c}-\omega_{v}$; here $\vartheta_{cv,\boldsymbol{k}}^{\left(1\right)}\left(t_{1}\right)=\left\langle cv,\boldsymbol{k}\right|V_{ext}\left(t_{1}\right)\left|0\right\rangle $
and $\vartheta_{cv,\boldsymbol{k}}^{\left(2\right)}\left(t_{1},t_{2}\right)=\left\langle cv,\boldsymbol{k}\right|V_{ext}\left(t_{1}\right)V_{ext}\left(t_{2}\right)\left|0\right\rangle $.
This allows us to compute the injection rate for the density $\left\langle M\right\rangle $
of a quantity associated with a single-particle operator ${\cal M}=\sum_{\boldsymbol{k}}a_{\alpha,\boldsymbol{k}}^{\dagger}M_{\alpha\beta,\boldsymbol{k}}a_{\beta,\boldsymbol{k}}$,
where $\alpha$ and $\beta$ are band indices. The expression for
$\left\langle \dot{M}\right\rangle $ is 
\begin{equation}
\frac{d}{dt}\left\langle M\right\rangle =\frac{1}{L^{D}}\sum_{cv,c^{\prime}v^{\prime},\boldsymbol{k}}\left(M_{c^{\prime}c,\boldsymbol{k}}\delta_{v^{\prime}v}-M_{v^{\prime}v,\boldsymbol{k}}\delta_{c^{\prime}c}\right)\frac{d}{dt}\left(c_{c^{\prime}v^{\prime},\boldsymbol{k}}^{\ast}c_{cv,\boldsymbol{k}}\right),
\end{equation}
where $L$ is the unidimensional length and $D$ is the spatial dimension
of the system.

Specifying the perturbation to be an incident laser field, using the minimal coupling Hamiltonian we have  \begin{equation}
\begin{array}{rl}
\vartheta_{cv,\boldsymbol{k}}\left(t_{1}\right)= & -\frac{e}{c}\boldsymbol{v}_{cv,\boldsymbol{k}}\cdot\boldsymbol{A}\left(t_{1}\right)e^{i\omega_{cv}t_{1}},\\
\vartheta_{cv,\boldsymbol{k}}^{\left(2\right)}\left(t_{1},t_{2}\right)= & \frac{e^{2}}{c^{2}}\underset{c^{\prime}}{\sum}\boldsymbol{v}_{cc^{\prime}}\cdot\boldsymbol{A}\left(t_{1}\right)e^{i\omega_{cc^{\prime}}t_{1}}\boldsymbol{v}_{c^{\prime}v}\cdot\boldsymbol{A}\left(t_{2}\right)e^{i\omega_{c^{\prime}v}t_{2}}\\
 & -\frac{e^{2}}{c^{2}}\underset{v^{\prime}}{\sum}\boldsymbol{v}_{v^{\prime}v}\cdot\boldsymbol{A}\left(t_{1}\right)e^{i\omega_{v^{\prime}v}t_{1}}\boldsymbol{v}_{cv^{\prime}}\cdot\boldsymbol{A}\left(t_{2}\right)e^{i\omega_{cv^{\prime}}t_{2}},
\end{array}
\end{equation}
where $e=-\left|e\right|$ is the charge of the electron, $\boldsymbol{v}$
is the velocity operator, and the vector potential is 
\begin{equation}
\boldsymbol{A}\left(t\right)=\sum_{\omega_{\alpha}}\boldsymbol{A}\left(\omega_{\alpha}\right)e^{-i\left(\omega_{\alpha}+i\epsilon\right)t},
\end{equation}
with $\omega_{\alpha}=\pm\omega,\pm2\omega$; here $\epsilon\to0^{+}$
describes the turning on of the field from $t=-\infty$.

From Eq. \eqref{eq:coefC} we can write $c_{cv,\boldsymbol{k}}\left(t\right)$
as 
\begin{equation}
c_{cv,\boldsymbol{k}}\left(t\right)=\sum_{n=1}^{4}K_{cv,\boldsymbol{k}}^{\left(n\right)}\left(\omega\right)\frac{e^{-i\left(n\omega+i\epsilon\right)t}}{n\omega-\omega_{cv}+i\epsilon},
\end{equation}
where 
\begin{equation}
\begin{array}{rl}
K_{cv,\boldsymbol{k}}^{\left(1\right)}\left(\omega\right)= & -\frac{e}{\hbar c}\boldsymbol{v}_{cv,\boldsymbol{k}}\cdot\boldsymbol{A}\left(\omega\right)+\\
 & +\frac{e^{2}}{\hbar^{2}c^{2}}\underset{v^{\prime}c^{\prime}}{\sum}\frac{\boldsymbol{v}_{cc^{\prime}}\cdot\boldsymbol{A}\left(-\omega\right)\boldsymbol{v}_{c^{\prime}v}\cdot\boldsymbol{A}\left(2\omega\right)}{2\omega-\omega_{c^{\prime}v}+i\epsilon}-\frac{\boldsymbol{v}_{v^{\prime}v}\cdot\boldsymbol{A}\left(-\omega\right)\boldsymbol{v}_{cv^{\prime}}\cdot\boldsymbol{A}\left(2\omega\right)}{2\omega-\omega_{cv^{\prime}}+i\epsilon}\\
 & +\frac{e^{2}}{\hbar^{2}c^{2}}\underset{v^{\prime}c^{\prime}}{\sum}\frac{\boldsymbol{v}_{cc^{\prime}}\cdot\boldsymbol{A}\left(2\omega\right)\boldsymbol{v}_{c^{\prime}v}\cdot\boldsymbol{A}\left(-\omega\right)}{-\omega-\omega_{c^{\prime}v}+i\epsilon}-\frac{\boldsymbol{v}_{v^{\prime}v}\cdot\boldsymbol{A}\left(2\omega\right)\boldsymbol{v}_{cv^{\prime}}\cdot\boldsymbol{A}\left(-\omega\right)}{-\omega-\omega_{cv^{\prime}}+i\epsilon},
\end{array}
\end{equation}
and 
\begin{equation}
\begin{array}{rl}
K_{cv,\boldsymbol{k}}^{\left(2\right)}\left(\omega\right)= & -\frac{e}{\hbar c}\boldsymbol{v}_{cv,\boldsymbol{k}}\cdot\boldsymbol{A}\left(2\omega\right)+\\
 & +\frac{e^{2}}{\hbar^{2}c^{2}}\underset{v^{\prime}c^{\prime}}{\sum}\frac{\boldsymbol{v}_{cc^{\prime}}\cdot\boldsymbol{A}\left(\omega\right)\boldsymbol{v}_{c^{\prime}v}\cdot\boldsymbol{A}\left(\omega\right)}{\omega-\omega_{c^{\prime}v}+i\epsilon}-\frac{\boldsymbol{v}_{v^{\prime}v}\cdot\boldsymbol{A}\left(\omega\right)\boldsymbol{v}_{cv^{\prime}}\cdot\boldsymbol{A}\left(\omega\right)}{\omega-\omega_{cv^{\prime}}+i\epsilon},
\end{array}
\end{equation}
with similar expressions for $K_{cv,\boldsymbol{k}}^{\left(3\right)}$
and $K_{cv,\boldsymbol{k}}^{\left(4\right)}$; the first only has
terms with products of field amplitudes $A\left(\pm\omega\right)A\left(\pm2\omega\right)$,
and the second only has terms with products $A\left(\pm2\omega\right)A\left(\pm2\omega\right)$.
Then we have 
\begin{equation}
\begin{array}{rl}
\frac{d}{dt}\left(c_{c^{\prime}v^{\prime},\boldsymbol{k}}^{\ast}c_{cv,\boldsymbol{k}}\right)_{t=0}= & \pi\underset{n}{\sum}K_{c^{\prime}v^{\prime},\boldsymbol{k}}^{\left(n\right)\ast}\left(\omega\right)K_{cv,\boldsymbol{k}}^{\left(n\right)}\left(\omega\right)\\
 & \cdot\left[\delta\left(n\omega-\omega_{cv,\boldsymbol{k}}\right)+\delta\left(n\omega-\omega_{c^{\prime}v^{\prime},\boldsymbol{k}}\right)\right].
\end{array}\label{eq:timeder}
\end{equation}
We assume that the amplitude $A\left(\pm2\omega\right)$ of the second
harmonic field is much smaller that the amplitude $A\left(\pm\omega\right)$
of the fundamental field. Since two-photon processes are much weaker
than one-photon processes, two-photon processes involving the second
harmonic are then neglected, and only $K_{cv,\boldsymbol{k}}^{\left(1\right)}\left(\omega\right)$
and $K_{cv,\boldsymbol{k}}^{\left(2\right)}\left(\omega\right)$ remain. Since $K_{cv,\boldsymbol{k}}^{\left(3\right)}\left(\omega\right)$
and $K_{cv,\boldsymbol{k}}^{\left(4\right)}\left(\omega\right)$ do not have linear terms in the field amplitude they do not support any interference process, but only two-photon absorption.

\begin{figure}[h]
\includegraphics[width=0.5\columnwidth]{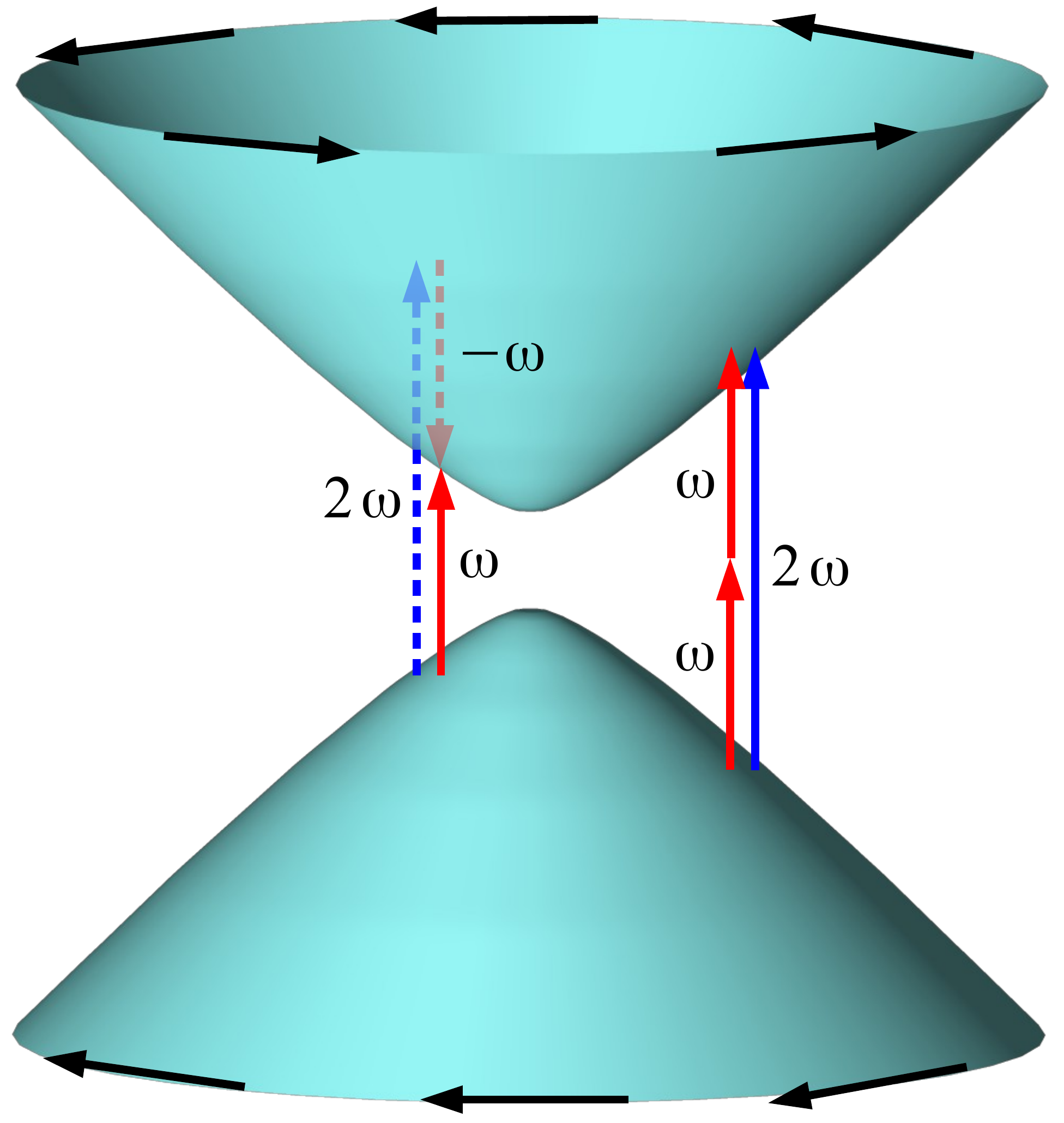} \caption[]{(Color online) One- and two-photon interference processes illustrates
on the helical Dirac cone. The one in the left has  energy $\hbar\omega$
corresponds to $(i)1$, while the other on the right has energy
$2\hbar\omega$ corresponds to $(i)2$. }

\label{fig:process} 
\end{figure}

Since all the $K_{cv,\boldsymbol{k}}^{\left(n\right)}\left(\omega\right)$
are multiplied by a $\delta$ function in Eq. \eqref{eq:timeder},
we can write 
\begin{equation}
\begin{array}{rl}
K_{cv,\boldsymbol{k}}^{\left(1\right)}\left(\omega\right)= & \frac{ie}{\hbar\omega} v_{cv,\boldsymbol{k}}^{b}E^{b}\left(\omega\right) \\
 & - {\cal W}_{cv,\boldsymbol{k}}^{bc}\left(2\omega,-\omega\right)E^{b}\left(-\omega\right)E^{c}\left(2\omega\right), \\
K_{cv,\boldsymbol{k}}^{\left(2\right)}\left(\omega\right)= & \frac{ie}{2\hbar\omega} v_{cv,\boldsymbol{k}}^{b}E^{b}\left(2\omega\right) \\ 
 & + {\cal W}_{cv,\boldsymbol{k}}^{bc}\left(\omega,\omega\right)E^{b}\left(\omega\right)E^{c}\left(\omega\right),
\end{array}
\end{equation}
where 
\begin{equation}
\begin{array}{rl}
\Omega_{cv,\boldsymbol{k}}^{bc}\left(\omega_{\alpha}\right)= & \frac{-e^{2}}{\hbar^{2}\omega^{2}}\underset{n}{\sum}\frac{v_{cn}^{b}v_{nv}^{c}}{\omega_{\alpha}-\omega_{nv}},\\
{\cal W}_{cv,\boldsymbol{k}}^{bc}\left(\omega_{\alpha},\omega_{\beta}\right)= & \Omega_{cv,\boldsymbol{k}}^{bc}\left(\omega_{\alpha}\right)+\Omega_{cv,\boldsymbol{k}}^{cb}\left(\omega_{\beta}\right),
\end{array}
\end{equation}
and $\boldsymbol{E}\left(t\right)=-c^{-1}\partial_{t}\boldsymbol{A}\left(t\right)$
was used, so $\boldsymbol{A}\left(n\omega\right)=-ic\left(n\omega\right)^{-1}\boldsymbol{E}\left(n\omega\right)$.
The injection rate for an operator ${\cal M}$ can then be decomposed
into contributions from one and two photons absorption processes with
an additional interference term $\left\langle \dot{M}\right\rangle =\left\langle \dot{M}_{1}\right\rangle +\left\langle \dot{M}_{2}\right\rangle +\left\langle \dot{M}_{i}\right\rangle $
where 
\begin{equation}
\begin{array}{rl}
\left\langle \dot{M}_{1}\right\rangle = & \underset{n=1,2}{\sum}\Lambda_{1}^{bc}\left(n\omega\right)E^{b}\left(-n\omega\right)E^{c}\left(n\omega\right),\\
\left\langle \dot{M}_{2}\right\rangle = & \Lambda_{2}^{bcde}\left(\omega\right)E^{b}\left(-\omega\right)E^{c}\left(-\omega\right)E^{d}\left(\omega\right)E^{e}\left(\omega\right),\\
\left\langle \dot{M}_{i}\right\rangle = & \underset{n=1,2}{\sum}\Lambda_{i\left(n\right)}^{bcd}\left(\omega\right)E^{b}\left(-\omega\right)E^{c}\left(-\omega\right)E^{d}\left(2\omega\right)+cc,
\end{array}
\end{equation}
with \begin{widetext} 
\begin{equation}
\begin{array}{rl}
\Lambda_{1}^{bc}\left(n\omega\right)= & \frac{\pi}{L^{D}}\underset{cv,c^{\prime}v^{\prime},\boldsymbol{k}}{\sum}\left(M_{c^{\prime}c,\boldsymbol{k}}\delta_{v^{\prime}v}-M_{v^{\prime}v,\boldsymbol{k}}\delta_{c^{\prime}c}\right)\Gamma_{1,c^{\prime}v^{\prime},cv}^{bc}\left(\boldsymbol{k},\omega\right)\left[\delta\left(n\omega-\omega_{cv,\boldsymbol{k}}\right)+\delta\left(n\omega-\omega_{c^{\prime}v^{\prime},\boldsymbol{k}}\right)\right],\\
\Lambda_{2}^{bcde}\left(\omega\right)= & \frac{\pi}{L^{D}}\underset{cv,c^{\prime}v^{\prime},\boldsymbol{k}}{\sum}\left(M_{c^{\prime}c,\boldsymbol{k}}\delta_{v^{\prime}v}-M_{v^{\prime}v,\boldsymbol{k}}\delta_{c^{\prime}c}\right)\Gamma_{2,c^{\prime}v^{\prime},cv}^{bcde}\left(\boldsymbol{k},\omega\right)\left[\delta\left(2\omega-\omega_{cv,\boldsymbol{k}}\right)+\delta\left(2\omega-\omega_{c^{\prime}v^{\prime},\boldsymbol{k}}\right)\right],\\
\Lambda_{i\left(n\right)}^{bcd}\left(\omega\right)= & \frac{\pi}{L^{D}}\underset{cv,c^{\prime}v^{\prime},\boldsymbol{k}}{\sum}\left(M_{c^{\prime}c,\boldsymbol{k}}\delta_{v^{\prime}v}-M_{v^{\prime}v,\boldsymbol{k}}\delta_{c^{\prime}c}\right)\Gamma_{i\left(n\right),c^{\prime}v^{\prime},cv}^{bcd}\left(\boldsymbol{k},\omega\right)\left[\delta\left(n\omega-\omega_{cv,\boldsymbol{k}}\right)+\delta\left(n\omega-\omega_{c^{\prime}v^{\prime},\boldsymbol{k}}\right)\right],
\end{array}\label{eq:coeffs}
\end{equation}
\end{widetext} and 
\begin{equation}
\begin{array}{rl}
\Gamma_{1,c^{\prime}v^{\prime},cv}^{bc}\left(\boldsymbol{k},\omega\right)= & \frac{e^{2}}{\hbar^{2}\omega^{2}}v_{v^{\prime}c^{\prime}}^{b}v_{cv}^{c},\\
\Gamma_{2,c^{\prime}v^{\prime},cv}^{bcde}\left(\boldsymbol{k},\omega\right)= & {\cal W}_{c^{\prime}v^{\prime},\boldsymbol{k}}^{bc}\left(\omega,\omega\right)^{\ast}{\cal W}_{cv,\boldsymbol{k}}^{de}\left(\omega,\omega\right),\\
\Gamma_{i\left(1\right),c^{\prime}v^{\prime},cv}^{bcd}\left(\boldsymbol{k},\omega\right)= & \frac{ie}{\hbar\omega}v_{v^{\prime}c^{\prime}}^{b}{\cal W}_{cv,\boldsymbol{k}}^{cd}\left(2\omega,-\omega\right),\\
\Gamma_{i\left(2\right),c^{\prime}v^{\prime},cv}^{bcd}\left(\boldsymbol{k},\omega\right)= & \frac{ie}{2\hbar\omega}{\cal W}_{cv,\boldsymbol{k}}^{bc}\left(\omega,\omega\right)^{\ast}v_{c^{\prime}v^{\prime}}^{d}.
\end{array}\label{eq:gammas}
\end{equation}
The $\Lambda_{1}^{bc}\left(\omega\right)$ and $\Lambda_{i\left(1\right)}^{bcd}\left(\omega\right)$
terms have usually been ignored in the literature, since they vanish
for systems with a gap where the first harmonic falls below the bandgap.
The two interference processes are shown in Fig. \ref{fig:process}.
The quantities for which the injection rates will be computed are
the densities associated with the carriers $\left\langle n\right\rangle $,
spin $\left\langle \boldsymbol{S}\right\rangle $, charge current
$\left\langle \boldsymbol{J}_{c}\right\rangle $, and spin current
$\left\langle \boldsymbol{J}_{S}\right\rangle $. We denote the response coefficients
associated with the quantities $\left\langle \dot{n}\right\rangle $,
$\left\langle \dot{\boldsymbol{S}}\right\rangle $, $\left\langle \dot{\boldsymbol{J}}_{c}\right\rangle $,
$\left\langle \dot{\boldsymbol{J}}_{S}\right\rangle $  respectively by 
$\xi$, $\zeta$, $\eta$, $\mu$.

\section{Two-band systems }

\label{sec:twoband}

Any Hermitian $2\times2$ matrix can be written as a linear combination
of Pauli matrices $\sigma$ and the identity $\sigma_{0}$. So a generic
Hamiltonian for two bands is ${\cal H}=\sum_{\boldsymbol{k}}a_{\alpha,\boldsymbol{k}}^{\dagger}H_{\alpha\beta,\boldsymbol{k}}a_{\beta,\boldsymbol{k}}$,
where $\alpha,\beta=1,2$ are band indices, and 
\begin{equation}
H_{\boldsymbol{k}}=\hbar\varpi_{\boldsymbol{k}}\sigma_{0}+\hbar\boldsymbol{d}_{\boldsymbol{k}}\cdot\boldsymbol{\sigma}=\hbar\left[\begin{array}{cc}
\varpi_{\boldsymbol{k}}+d_{\boldsymbol{k}}^{z} & d_{\boldsymbol{k}}^{x}-id_{\boldsymbol{k}}^{y}\\
d_{\boldsymbol{k}}^{x}+id_{\boldsymbol{k}}^{y} & \varpi_{\boldsymbol{k}}-d_{\boldsymbol{k}}^{z}
\end{array}\right]\label{eq:H2d}
\end{equation}
denotes the Hamiltonian at each lattice momentum $\boldsymbol{k}$.
The eigenenergies are $E_{\boldsymbol{k}\pm}=\hbar\left(\varpi_{\boldsymbol{k}}\pm d_{\boldsymbol{k}}\right)$
where $d_{\boldsymbol{k}}=\left|\boldsymbol{d}_{\boldsymbol{k}}\right|$,
with $\left(+\right)=c$ and $\left(-\right)=v$ representing the
conduction and valence bands respectively, so $\omega_{cv,\boldsymbol{k}}=2d_{\boldsymbol{k}}$.
The eigenstates satisfy $\hat{\boldsymbol{d}}_{\boldsymbol{k}}\cdot\boldsymbol{\sigma}\psi_{\boldsymbol{k}\pm}=\pm\psi_{\boldsymbol{k}\pm}$,
so when $\hat{\boldsymbol{d}}_{\boldsymbol{k}}\cdot\boldsymbol{\sigma}$
is diagonalized it is represented by $\sigma_{z}$, and there is a
unitary matrix $U_{\boldsymbol{k}}$ that performs the change of basis,
$\hat{\boldsymbol{d}}_{\boldsymbol{k}}\cdot\boldsymbol{\sigma}=U_{\boldsymbol{k}}\sigma_{z}U_{\boldsymbol{k}}^{\dagger}$.
Because $SU\left(2\right)$ and $SO\left(3\right)$ have the same
algebra, we can write $U_{\boldsymbol{k}}\sigma_{z}U_{\boldsymbol{k}}^{\dagger}=\left({\cal R}_{\boldsymbol{k}}\hat{\boldsymbol{z}}\right)\cdot\boldsymbol{\sigma}$,
where ${\cal R}_{\boldsymbol{k}}$ represents a rotation around the
axis $\hat{\boldsymbol{n}}_{\boldsymbol{k}}$ by an angle $\phi_{\boldsymbol{k}}$,
so $U_{\boldsymbol{k}}=\exp\left(-i\frac{\phi_{\boldsymbol{k}}}{2}\hat{\boldsymbol{n}}_{\boldsymbol{k}}\cdot\boldsymbol{\sigma}\right)$;
we put $\hat{\boldsymbol{n}}_{\boldsymbol{k}}=\hat{\boldsymbol{z}}\times\hat{\boldsymbol{d}}_{\boldsymbol{k}}/\left|\hat{\boldsymbol{z}}\times\hat{\boldsymbol{d}}_{\boldsymbol{k}}\right|$
and $\cos\phi_{\boldsymbol{k}}=\hat{z}\cdot\hat{\boldsymbol{d}}_{\boldsymbol{k}}$.
The triad $\left(\hat{\boldsymbol{n}}_{\boldsymbol{k}},\hat{\boldsymbol{d}}_{\boldsymbol{k}},\hat{\boldsymbol{n}}_{\boldsymbol{k}}\times\hat{\boldsymbol{d}}_{\boldsymbol{k}}\right)$
forms an orthonormal basis, so an arbitrary operator can be written
as 
\begin{equation}
\begin{array}{rl}
\hat{\boldsymbol{w}}\cdot\boldsymbol{\sigma}= & \left(\hat{\boldsymbol{n}}_{\boldsymbol{k}}\cdot\hat{\boldsymbol{w}}\right)\hat{\boldsymbol{n}}_{\boldsymbol{k}}\cdot\boldsymbol{\sigma}+\left(\hat{\boldsymbol{d}}_{\boldsymbol{k}}\cdot\hat{\boldsymbol{w}}\right)\hat{\boldsymbol{d}}_{\boldsymbol{k}}\cdot\boldsymbol{\sigma}\\
 & +\left[\left(\hat{\boldsymbol{n}}_{\boldsymbol{k}}\times\hat{\boldsymbol{d}}_{\boldsymbol{k}}\right)\cdot\hat{\boldsymbol{w}}\right]\left(\hat{\boldsymbol{n}}_{\boldsymbol{k}}\times\hat{\boldsymbol{d}}_{\boldsymbol{k}}\right)\cdot\boldsymbol{\sigma},
\end{array}
\end{equation}
and in the basis of eigenvectors 
\begin{equation}
\begin{array}{rl}
U_{\boldsymbol{k}}^{\dagger}\left(\hat{\boldsymbol{w}}\cdot\boldsymbol{\sigma}\right)U_{\boldsymbol{k}}= & \left(\hat{\boldsymbol{n}}_{\boldsymbol{k}}\cdot\hat{\boldsymbol{w}}\right)\hat{\boldsymbol{n}}_{\boldsymbol{k}}\cdot\boldsymbol{\sigma}+\left(\hat{\boldsymbol{d}}_{\boldsymbol{k}}\cdot\hat{\boldsymbol{w}}\right)\hat{\boldsymbol{z}}\cdot\boldsymbol{\sigma}\\
 & +\left[\left(\hat{\boldsymbol{n}}_{\boldsymbol{k}}\times\hat{\boldsymbol{d}}_{\boldsymbol{k}}\right)\cdot\hat{\boldsymbol{w}}\right]\left(\hat{\boldsymbol{n}}_{\boldsymbol{k}}\times\hat{\boldsymbol{z}}\right)\cdot\boldsymbol{\sigma},
\end{array}\label{eq:eigen}
\end{equation}
which allows any operator to be expressed simply.

\subsubsection*{Operators }

The quantities of interest are the densities of injected carriers
$\left\langle n\right\rangle $, spin $\left\langle \boldsymbol{S}\right\rangle $,
charge current $\left\langle \boldsymbol{J}_{c}\right\rangle $ and
spin current $\left\langle \boldsymbol{J}_{S}\right\rangle $, which
are computed below.

We keep track of the injected carriers by computing the density of
electrons injected into the conduction band. The corresponding number
operator has matrix elements $n_{cc}=1$ and $n_{vv}=0$.

We suppose that the components of the spin operator are given by $S^{a}=\frac{\hbar}{2}\hat{\boldsymbol{a}}\cdot\boldsymbol{\sigma}$, and decompose $\hat{\boldsymbol{a}}\cdot\boldsymbol{\sigma}$
according to Eq. \eqref{eq:eigen}, so 
\begin{equation}
\begin{array}{rl}
S_{cc}^{a}= & \frac{\hbar}{2}{\rm Tr}\left[S^{a}\left(I+\sigma_{z}\right)\right]=\frac{\hbar}{2}\hat{\boldsymbol{d}}_{\boldsymbol{k}}\cdot\hat{\boldsymbol{a}},\\
S_{vv}^{a}= & \frac{\hbar}{2}{\rm Tr}\left[S^{a}\left(I-\sigma_{z}\right)\right]=-\frac{\hbar}{2}\hat{\boldsymbol{d}}_{\boldsymbol{k}}\cdot\hat{\boldsymbol{a}}
\end{array}\label{eq:compspin}
\end{equation}
are the matrix elements needed. Note that even though $S_{cc}^{a}$
and $S_{vv}^{a}$ are matrix elements of the spin operator in the
basis of eigenstates, they are being expressed in terms of the parameters
of the Hamiltonian in its non-diagonal form of Eq. \eqref{eq:H2d}.

The matrix associated with the velocity operator is %
\footnote{Here we are still using a discrete momentum basis, the derivative
can be obtained from the extension of the function $H_{\boldsymbol{k}}$
to continuum momenta and then restricting $\partial_{k^{a}}H_{\boldsymbol{k}}$
back to discrete momentum space. %
} 
\begin{equation}
v_{\boldsymbol{k}}^{a}=\frac{1}{\hbar}\partial_{k^{a}}H_{\boldsymbol{k}}=\partial_{k^{a}}\varpi_{\boldsymbol{k}}\sigma_{0}+\partial_{k^{a}}\boldsymbol{d}_{\boldsymbol{k}}\cdot\boldsymbol{\sigma},
\end{equation}
 and decomposing it according to Eq. \eqref{eq:eigen} gives 
\begin{equation}
\begin{array}{rl}
v_{\boldsymbol{k}}^{a}= & \partial_{k^{a}}\varpi_{\boldsymbol{k}}\sigma_{0}+\partial_{k^{a}}d_{\boldsymbol{k}}\sigma_{z}+d_{\boldsymbol{k}}\left(\hat{\boldsymbol{n}}_{\boldsymbol{k}}\cdot\partial_{k^{a}}\hat{\boldsymbol{d}}_{\boldsymbol{k}}\right)\hat{\boldsymbol{n}}_{\boldsymbol{k}}\cdot\sigma\\
 & +d_{\boldsymbol{k}}\left[\left(\hat{\boldsymbol{n}}_{\boldsymbol{k}}\times\hat{\boldsymbol{d}}_{\boldsymbol{k}}\right)\cdot\partial_{k^{a}}\hat{\boldsymbol{d}}_{\boldsymbol{k}}\right]\left(\hat{\boldsymbol{n}}_{\boldsymbol{k}}\times\hat{\boldsymbol{z}}\right)\cdot\sigma.
\end{array}
\end{equation}
The velocity matrix elements are 
\begin{equation}
\begin{array}{rl}
v_{cc}^{a}= & \partial_{k^{a}}\varpi_{\boldsymbol{k}}+\partial_{k^{a}}d_{\boldsymbol{k}},\\
v_{vv}^{a}= & \partial_{k^{a}}\varpi_{\boldsymbol{k}}-\partial_{k^{a}}d_{\boldsymbol{k}},\\
v_{cv}^{a}= & d_{\boldsymbol{k}}\left(\hat{\boldsymbol{n}}_{\boldsymbol{k}}+i\hat{\boldsymbol{n}}_{\boldsymbol{k}}\times\hat{\boldsymbol{d}}_{\boldsymbol{k}}\right)\cdot\left(\partial_{k^{a}}\hat{\boldsymbol{d}}_{\boldsymbol{k}}\right)\hat{\boldsymbol{n}}_{\boldsymbol{k}}\cdot\left(\hat{\boldsymbol{x}}-i\hat{\boldsymbol{y}}\right),\\
v_{vc}^{a}= & d_{\boldsymbol{k}}\left(\hat{\boldsymbol{n}}_{\boldsymbol{k}}-i\hat{\boldsymbol{n}}_{\boldsymbol{k}}\times\hat{\boldsymbol{d}}_{\boldsymbol{k}}\right)\cdot\left(\partial_{k^{a}}\hat{\boldsymbol{d}}_{\boldsymbol{k}}\right)\hat{\boldsymbol{n}}_{\boldsymbol{k}}\cdot\left(\hat{\boldsymbol{x}}+i\hat{\boldsymbol{y}}\right).
\end{array}\label{eq:compveloc}
\end{equation}
It is also necessary to compute products of two velocity matrix elements,
\begin{equation}
v_{cv}^{a}v_{vc}^{b}=d_{\boldsymbol{k}}^{2}\left[\partial_{k^{a}}\hat{\boldsymbol{d}}_{\boldsymbol{k}}\cdot\partial_{k^{b}}\hat{\boldsymbol{d}}_{\boldsymbol{k}}+i\hat{\boldsymbol{d}}_{\boldsymbol{k}}\cdot\left(\partial_{k^{a}}\hat{\boldsymbol{d}}_{\boldsymbol{k}}\times\partial_{k^{b}}\hat{\boldsymbol{d}}_{\boldsymbol{k}}\right)\right].\label{eq:velvel}
\end{equation}
The second term above is the Berry curvature; we can track the contributions
to optical properties that depend on it. The charge current is expressed
in terms of the velocity operator by $\boldsymbol{J}=e\boldsymbol{v}$.

We define the spin current operator as $J_{S,\boldsymbol{k}}^{ab}=\frac{1}{2}\left(S^{a}v_{\boldsymbol{k}}^{b}+v_{\boldsymbol{k}}^{b}S^{a}\right)$
so for a system where $S^{a}=\frac{\hbar}{2}\hat{\boldsymbol{a}}\cdot\boldsymbol{\sigma}$
we have 
\begin{equation}
\begin{array}{rl}
U_{\boldsymbol{k}}^{\dagger}J_{S,\boldsymbol{k}}^{ab}U_{\boldsymbol{k}}= & \frac{\hbar}{2}\left[\hat{\boldsymbol{a}}\cdot\hat{\boldsymbol{d}}_{\boldsymbol{k}}\left(\partial_{k^{b}}d_{\boldsymbol{k}}\right)+d_{\boldsymbol{k}}\left(\hat{\boldsymbol{a}}\cdot\partial_{k^{b}}\hat{\boldsymbol{d}}_{\boldsymbol{k}}\right)\right]+\\
 & +\left(\partial_{k^{b}}\varpi_{\boldsymbol{k}}\right)U_{\boldsymbol{k}}^{\dagger}S^{a}U_{\boldsymbol{k}},
\end{array}
\end{equation}
and the components are 
\begin{equation}
\begin{array}{rl}
J_{S,cc}^{ab}= & \hbar \frac{\hat{\boldsymbol{a}}\cdot\hat{\boldsymbol{d}}_{\boldsymbol{k}}\left(\partial_{k^{b}}d_{\boldsymbol{k}}\right)+d_{\boldsymbol{k}}\hat{\boldsymbol{a}}\cdot\partial_{k^{b}}\hat{\boldsymbol{d}}_{\boldsymbol{k}}+\left(\partial_{k^{b}}\varpi_{\boldsymbol{k}}\right)\hat{\boldsymbol{a}}\cdot\hat{\boldsymbol{d}}_{\boldsymbol{k}}}{2},\\
J_{S,vv}^{ab}= & \hbar\frac{\hat{\boldsymbol{a}}\cdot\hat{\boldsymbol{d}}_{\boldsymbol{k}}\left(\partial_{k^{b}}d_{\boldsymbol{k}}\right)+d_{\boldsymbol{k}}\hat{\boldsymbol{a}}\cdot\partial_{k^{b}}\hat{\boldsymbol{d}}_{\boldsymbol{k}}-\left(\partial_{k^{b}}\varpi_{\boldsymbol{k}}\right)\hat{\boldsymbol{a}}\cdot\hat{\boldsymbol{d}}_{\boldsymbol{k}}}{2},
\end{array}\label{eq:compspincurr}
\end{equation}
which completes the list of necessary matrix elements.

\subsubsection*{Optical injection coefficients }

The quantities necessary for computing the injection rates are readily
obtained from the equations above, giving 
\begin{equation}
\begin{array}{rl}
\Omega_{cv,\boldsymbol{k}}^{bc}\left(\omega_{\alpha}\right)= & \frac{-e^{2}}{\hbar^{2}\omega^{2}}\left(\frac{v_{cc}^{b}v_{cv}^{c}}{\omega_{\alpha}-\omega_{cv}}+\frac{v_{cv}^{b}v_{vv}^{c}}{\omega_{\alpha}}\right),\\
{\cal W}_{cv,\boldsymbol{k}}^{bc}\left(\omega_{\alpha},\omega_{\beta}\right)= & \frac{e^{2}}{\hbar^{2}\omega^{2}}\left[\frac{v_{cv}^{b}\partial_{k^{c}}d_{\boldsymbol{k}}}{\omega_{\alpha}}+\frac{v_{cv}^{c}\partial_{k^{b}}d_{\boldsymbol{k}}}{\omega_{\beta}}\right],
\end{array}
\end{equation}
where we have used the fact that $\omega_{\alpha}+\omega_{\beta}=\omega_{cv}$.
Therefore 
\begin{equation}
\begin{array}{rl}
\Gamma_{1,cv}^{bc}\left(\boldsymbol{k},\omega\right)= & \frac{e^{2}d_{\boldsymbol{k}}^{2}\left[\partial_{k^{b}}\hat{\boldsymbol{d}}_{\boldsymbol{k}}\cdot\partial_{k^{c}}\hat{\boldsymbol{d}}_{\boldsymbol{k}}-i\hat{\boldsymbol{d}}_{\boldsymbol{k}}\cdot\left(\partial_{k^{b}}\hat{\boldsymbol{d}}_{\boldsymbol{k}}\times\partial_{k^{c}}\hat{\boldsymbol{d}}_{\boldsymbol{k}}\right)\right]}{\hbar^{2}\omega^{2}},\\
\Gamma_{2,cv}^{bcde}\left(\boldsymbol{k},\omega\right)= & \frac{e^{4}}{\hbar^{4}\omega^{4}}\left[\frac{v_{cv}^{d}\partial_{k^{e}}d_{\boldsymbol{k}}+v_{cv}^{e}\partial_{k^{d}}d_{\boldsymbol{k}}}{\omega}\right]\left[\frac{v_{vc}^{b}\partial_{k^{c}}d_{\boldsymbol{k}}+v_{vc}^{c}\partial_{k^{b}}d_{\boldsymbol{k}}}{\omega}\right],
\end{array}\label{eq:gammas2dA}
\end{equation}
and 
\begin{equation}
\begin{array}{rl}
\Gamma_{i\left(1\right),cv}^{bcd}\left(\boldsymbol{k},\omega\right)= & \frac{ie^{3}}{\hbar^{3}\omega^{3}}\left[\frac{v_{cv}^{c}v_{vc}^{b}\partial_{k^{d}}d_{\boldsymbol{k}}}{2\omega}-\frac{v_{cv}^{d}v_{vc}^{b}\partial_{k^{c}}d_{\boldsymbol{k}}}{\omega}\right],\\
\Gamma_{i\left(2\right),cv}^{bcd}\left(\boldsymbol{k},\omega\right)= & \frac{ie^{3}}{2\hbar^{3}\omega^{3}}\left[\frac{v_{cv}^{d}v_{vc}^{b}\partial_{k^{c}}d_{\boldsymbol{k}}+v_{cv}^{d}v_{vc}^{c}\partial_{k^{b}}d_{\boldsymbol{k}}}{\omega}\right].
\end{array}\label{eq:gammas2dB}
\end{equation}
For a two-band system there is only one valence and one conduction
band, $c^{\prime}=c$ and $v^{\prime}=v$, so Eqs. \eqref{eq:coeffs}
and \eqref{eq:gammas} become simpler; in the continuum limit the
momentum sums are expressed by 
\begin{equation}
\begin{array}{rl}
\Lambda_{1}^{bc}\left(n\omega\right)= & \underset{\omega_{cv}=n\omega}{\int}\frac{\left(dk\right)^{D-1}}{\left(2\pi\right)^{D-1}}\frac{\left(M_{cc,\boldsymbol{k}}-M_{vv,\boldsymbol{k}}\right)\Gamma_{1,cv}^{bc}\left(\boldsymbol{k},n\omega\right)}{\left|\boldsymbol{\nabla}_{\boldsymbol{k}}\omega_{cv}\right|},\\
\Lambda_{2}^{bcde}\left(\omega\right)= & \underset{\omega_{cv}=2\omega}{\int}\frac{\left(dk\right)^{D-1}}{\left(2\pi\right)^{D-1}}\frac{\left(M_{cc,\boldsymbol{k}}-M_{vv,\boldsymbol{k}}\right)\Gamma_{2,cv}^{bcde}\left(\boldsymbol{k},\omega\right)}{\left|\boldsymbol{\nabla}_{\boldsymbol{k}}\omega_{cv}\right|},\\
\Lambda_{i\left(n\right)}^{bcd}\left(\omega\right)= & \underset{\omega_{cv}=n\omega}{\int}\frac{\left(dk\right)^{D-1}}{\left(2\pi\right)^{D-1}}\frac{\left(M_{cc,\boldsymbol{k}}-M_{vv,\boldsymbol{k}}\right)\Gamma_{i,cv,\left(n\right)}^{bcd}\left(\boldsymbol{k},\omega\right)}{\left|\boldsymbol{\nabla}_{\boldsymbol{k}}\omega_{cv}\right|},
\end{array}\label{eq:coeff2band}
\end{equation}
where each $D-1$ dimensional integral is over the region specified
by the energy matching condition $\omega_{cv}=n\omega$.

\section{Topological insulators }

\label{sec:topins}

When the photon energy is smaller than the bulk band gap of a topological
insulators, only the protected states localized on their surfaces
will contribute to the optical absorption and injection.

The standard effective model obtained from a $\boldsymbol{k}\cdot\boldsymbol{p}$
approximation for the materials in the family of BiSeTe results in
a four-band Hamiltonian \cite{liu10} 
\begin{equation}
H_{\boldsymbol{k}}=\varepsilon_{\boldsymbol{k}}^{0}I_{4\times4}+\hbar\left[\begin{array}{cccc}
m_{\boldsymbol{k}} & A_{1}k_{z} & 0 & A_{2}k_{-}\\
A_{1}k_{z} & -m_{\boldsymbol{k}} & A_{2}k_{-} & 0\\
0 & A_{2}k_{+} & m_{\boldsymbol{k}} & -A_{1}k_{z}\\
A_{2}k_{+} & 0 & -A_{1}k_{z} & -m_{\boldsymbol{k}}
\end{array}\right],\label{eq:tiHbulk}
\end{equation}
where $k_{\pm}=k_{x}\pm ik_{y}$, $\varepsilon_{k}^{0}=C+D_{1}k_{z}^{2}+D_{2}k_{\perp}^{2}$
and $m_{\boldsymbol{k}}=m-B_{1}k_{z}^{2}-B_{2}k_{\perp}^{2}$ with
the constants depending on the material.

In order to consider interfaces along the $\hat{\boldsymbol{z}}$
direction one can write the Hamiltonian in a separated form $H_{\boldsymbol{k}}=H_{\boldsymbol{k}}^{\perp}+H_{\boldsymbol{k}}^{z}$,
and in the limit of $\boldsymbol{k}_{\perp}\to0$ the transverse part
$H_{\boldsymbol{k}}^{\perp}$ can be neglected. Also, $H_{\boldsymbol{k}}^{z}$
is block diagonal and separates into two sectors according to the
spin of the electrons. The boundary conditions for surface states
then lead to only one solution for each sector, giving two independent
states. Next, the $4\times4$ bulk Hamiltonian with lattice momentum
near the $\Gamma$ point is projected on the subspace spanned by the
two independent surface states, and an effective Hamiltonian is obtained
for the surface states, 
\begin{equation}
H_{\boldsymbol{k}_{\perp}}=\hbar\left(C_{0}+D_{0}k^{2}\right)\sigma_{0}-\hbar A_{0}\left(\hat{\boldsymbol{z}}\times\boldsymbol{k}\right)\cdot\boldsymbol{\sigma},\label{eq:tiHsurf}
\end{equation}
which is valid in the limit of large slab thickness \cite{lu10} and
including terms up to the second order in $\boldsymbol{k}$. The parameters
$C_{0}$, $D_{0}$ and $A_{0}$ can be determined in terms of the
bulk parameters appearing in Eq. \eqref{eq:tiHbulk}.

In order to identify the most basic features of coherent control optical
injection in topological insulators we will compute the injection
rates starting from the Hamiltonian of Eq. \eqref{eq:tiHsurf}. However,
in order to keep track of how the Berry curvature effects the optical
response we keep a $\sigma_{z}$ mass term that would correspond,
for instance, to an external magnetic field along the $\hat{\boldsymbol{z}}$
direction. The 2D Hamiltonian we consider is then 
\begin{equation}
H_{\boldsymbol{k}_{\perp}}=\hbar\left(C_{0}+D_{0}k^{2}\right)\sigma_{0}-\hbar A_{0}\left(\hat{\boldsymbol{z}}\times\boldsymbol{k}\right)\cdot\boldsymbol{\sigma}+\hbar\Delta_{m}\sigma_{z},
\end{equation}
which corresponds to Eq. \eqref{eq:H2d} with $\varpi_{\boldsymbol{k}}=C_{0}-Dk^{2}$
and $\boldsymbol{d}_{\boldsymbol{k}}=A_{0}\left(\hat{\boldsymbol{z}}\times\boldsymbol{k}\right)+\Delta_{m}\hat{\boldsymbol{z}}$
so $\hat{\boldsymbol{n}}_{\boldsymbol{k}}=-\hat{\boldsymbol{k}}$
and 
\begin{equation}
\begin{array}{rl}
\partial_{k^{b}}d_{\boldsymbol{k}}= & \frac{A_{0}^{2}k^{b}}{d_{\boldsymbol{k}}},\\
\partial_{k^{b}}\hat{\boldsymbol{d}}_{\boldsymbol{k}}= & \frac{-A_{0}\left(\hat{\boldsymbol{z}}\times\hat{\boldsymbol{b}}\right)}{d_{\boldsymbol{k}}}-\frac{A_{0}^{2}k^{b}\hat{\boldsymbol{d}}_{\boldsymbol{k}}}{d_{\boldsymbol{k}}^{2}}.
\end{array}
\end{equation}
This allows us to compute the optical injection coefficients. Since
in the basis of Eq. \eqref{eq:tiHsurf} the spin operator is represented
by $S^{a}=\frac{\hbar}{2}\hat{\boldsymbol{a}}\cdot\boldsymbol{\sigma}$,
from Eq. \eqref{eq:compspin}, Eq. \eqref{eq:compveloc} and Eq. \eqref{eq:compspincurr}
we can identify the matrix elements of the operators of interest 
\begin{equation}
\begin{array}{rl}
S_{cc}^{a}-S_{vv}^{a}= & \hbar\hat{\boldsymbol{a}}\cdot\hat{\boldsymbol{d}}_{\boldsymbol{k}}=\frac{\hbar A_{0}k^{z\times a}+\hbar\Delta_{m}\hat{\boldsymbol{z}}\cdot\hat{\boldsymbol{a}}}{d_{\boldsymbol{k}}},\\
v_{cc}^{a}-v_{vv}^{a}= & 2\partial_{k^{a}}d_{\boldsymbol{k}}=\frac{2A_{0}^{2}k^{a}}{d_{\boldsymbol{k}}},\\
J_{S,cc}^{ab}-J_{S,vv}^{ab}= & \hbar\left(\partial_{k^{b}}\varepsilon_{k}\right)\hat{\boldsymbol{a}}\cdot\hat{\boldsymbol{d}}_{\boldsymbol{k}}=\frac{2\hbar D_{0}k^{b}\left[A_{0}k^{z\times a}+\Delta_{m}\hat{\boldsymbol{z}}\cdot\hat{\boldsymbol{a}}\right]}{d_{\boldsymbol{k}}}.
\end{array}\label{eq:ticomps}
\end{equation}
They satisfy the relations 
\begin{equation}
\begin{array}{rl}
S_{cc}^{z}-S_{vv}^{z}= & \frac{\hbar\Delta_{m}}{d_{\boldsymbol{k}}}=\frac{\hbar\Delta_{m}}{d_{\boldsymbol{k}}}\left(n_{cc}-n_{vv}\right),\vspace{2pt}\\
\boldsymbol{v}= & -\frac{2A_{0}\hat{\boldsymbol{z}}\times\boldsymbol{S}}{\hbar},\vspace{2pt}\\
J_{S,cc}^{zb}-J_{S,vv}^{zb}= & \frac{2\hbar D_{0}k^{b}\Delta_{m}}{d_{\boldsymbol{k}}}=\frac{\hbar D_{0}\Delta_{m}}{A_{0}^{2}}\left(v_{cc}^{b}-v_{vv}^{b}\right),
\end{array}\label{eq:relations}
\end{equation}
where the second equation is the identity explored by Raghu et al.
{[}\onlinecite{raghu10}{]}; the first states that the $\hat{\boldsymbol{z}}$
component of the spin density $S^{z}$ merely corresponds to the spin
polarization of the injected carriers; and the third identifies the
$\hat{\boldsymbol{z}}$ component of the spin current $J_{S}^{z}$
as entirely due to the spin polarization of the charge current. Both
spin density and current are non-zero only in the presence of the
$\sigma_{z}$ mass $\Delta_{m}$. It should be noted that the spin
current is typically not a conserved quantity, and indeed it is not
conserved at the surface of topological insulators. Nevertheless,
we still compute its optical injection rate because depending on the
experimental technique, the spin separation to which it leads might
be detected (or tunneled to another material) before the spins relax
\cite{driel06prl,driel06ssc}.

Eq. \eqref{eq:velvel} is then 
\begin{equation}
v_{cv}^{b}v_{vc}^{c}=A_{0}^{2}\hat{\boldsymbol{b}}\cdot\hat{\boldsymbol{c}}-\frac{A_{0}^{4}k^{b}k^{c}}{d_{\boldsymbol{k}}^{2}}+i\frac{A_{0}^{2}\Delta_{m}\hat{\boldsymbol{z}}\cdot\left(\hat{\boldsymbol{b}}\times\hat{\boldsymbol{c}}\right)}{d_{\boldsymbol{k}}}\label{eq:tivelvel}
\end{equation}
which allows us to compute the $\Gamma$ coefficients of Eq. \eqref{eq:gammas2dA}
and Eq. \eqref{eq:gammas2dB}, giving 
\begin{equation}
\begin{array}{rl}
\Gamma_{1,cv}^{bc}\left(\boldsymbol{k},\omega\right)= & \frac{e^{2}}{\hbar^{2}\omega^{2}}v_{cv}^{c}v_{vc}^{b},\\
\Gamma_{2,cv}^{bcde}\left(\boldsymbol{k},\omega\right)= & \frac{e^{4}A_{0}^{4}}{\hbar^{4}\omega^{4}}\left[\frac{k^{e}v_{cv}^{d}+k^{d}v_{cv}^{e}}{\omega d_{\boldsymbol{k}}}\right]\left[\frac{k^{c}v_{vc}^{b}+k^{b}v_{vc}^{c}}{\omega d_{\boldsymbol{k}}}\right],
\end{array}\label{eq:tigamma12}
\end{equation}
and 
\begin{equation}
\begin{array}{rl}
\Gamma_{i\left(1\right),cv}^{bcd}\left(\boldsymbol{k},\omega\right)= & \frac{ie^{3}A_{0}^{2}}{2\hbar^{3}\omega^{3}}\left[\frac{k^{d}v_{cv}^{c}v_{vc}^{b}-2k^{c}v_{cv}^{d}v_{vc}^{b}}{\omega d_{\boldsymbol{k}}}\right],\\
\Gamma_{i\left(2\right),cv}^{bcd}\left(\boldsymbol{k},\omega\right)= & \frac{ie^{3}A_{0}^{2}}{2\hbar^{3}\omega^{3}}\left[\frac{k^{c}v_{cv}^{d}v_{vc}^{b}+k^{b}v_{cv}^{d}v_{vc}^{c}}{\omega d_{\boldsymbol{k}}}\right].
\end{array}\label{eq:tigammaint}
\end{equation}
The optical injection coefficients can now be computed.

The only relevant states for our calculations are localized on the
surface, so the momentum integrals are all two dimensional. Since
$\omega_{cv,\boldsymbol{k}}$ does not depend on the direction of
the momentum, when the integrals are performed in polar coordinates
$\left(k,\theta\right)$, the $k$ integral simply deals with the
delta function setting $2d_{\boldsymbol{k}}=\omega$ or $2d_{\boldsymbol{k}}=2\omega$
depending on the term. So Eq. \eqref{eq:coeff2band} becomes 
\begin{equation}
\begin{array}{rl}
\Lambda_{1}^{bc}\left(n\omega\right)= & \left.\int\frac{d\theta}{2\pi}\frac{d_{\boldsymbol{k}}\left(M_{cc,\boldsymbol{k}}-M_{vv,\boldsymbol{k}}\right)\Gamma_{1,cv}^{bc}\left(\boldsymbol{k},n\omega\right)}{2A_{0}^{2}}\right|_{d_{\boldsymbol{k}}=\frac{n\omega}{2}},\\
\Lambda_{2}^{bcde}\left(\omega\right)= & \left.\int\frac{d\theta}{2\pi}\frac{d_{\boldsymbol{k}}\left(M_{cc,\boldsymbol{k}}-M_{vv,\boldsymbol{k}}\right)\Gamma_{2,cv}^{bcde}\left(\boldsymbol{k},\omega\right)}{2A_{0}^{2}}\right|_{d_{\boldsymbol{k}}=\omega},\\
\Lambda_{i\left(n\right)}^{bcd}\left(\omega\right)= & \left.\int\frac{d\theta}{2\pi}\frac{d_{\boldsymbol{k}}\left(M_{cc,\boldsymbol{k}}-M_{vv,\boldsymbol{k}}\right)\Gamma_{i,cv,\left(n\right)}^{bcd}\left(\boldsymbol{k},\omega\right)}{2A_{0}^{2}}\right|_{d_{\boldsymbol{k}}=\frac{n\omega}{2}}.
\end{array}\label{eq:coeffinal}
\end{equation}
 The expressions for the various coefficients that follow from these
expressions are the main results of this paper, and are detailed in
Appendix \ref{app:coeffs}.

\section{Results }

\label{sec:results}

For the system we are considering, one- and two-photon absorption
processes inject scalar quantities while interference processes inject
vectorial ones. We confirm within our model that carriers are injected
by one- and two-photon absorption processes, but not from the interference
between them. Conversely, charge current is injected solely from the
interference processes and not from the one- and two-photon absorption
processes. However, there are additional peculiarities for the spin
density and spin current injection.

Due to the relations \eqref{eq:relations}, the in-plane spin density
follows the charge current injection, stemming only from the interference
processes; the out-of-plane spin density only has contributions from
the one- and two-photon absorption processes. It simply corresponds
to the spin polarization of the injected carriers, which is proportional
to the $\sigma_{z}$ mass term in the Hamiltonian.

A similar situation holds for the spin current. The spin current of
the $\hat{\boldsymbol{z}}$ component of spin follows the charge current
and simply amounts to the net spin polarization of the carriers of
the current; it is obtained from the interference terms. On the other
hand, the in-plane spin current is a result of the Dirac cone with
chiral spins; it does not require a net spin polarization generated by a $\sigma_{z}$ mass term. It is obtained from one- and two-photon
absorption and has no contribution from interference processes.

Below we present the injection rates for the quantities of interest,
considering linear and circular polarizations. In Appendix \ref{app:formula}
we show the general expressions for the optical injection coefficients,
and in Appendix \ref{app:coeffs} we present the explicit form of
the coefficients related to linear and circular polarizations of the
incident light, which are referred to below.

The values of the parameters $A_{0}$, $D_{0}$ and $\Delta_{m}=\frac{\mu_{B}g}{2\hbar}B$
used for the plots or specific estimates are given in Table \ref{tab:param};
they correspond to the parameters of $Bi_{2}Te_{3}$ for an applied
magnetic field around $10T$. \cite{liu10}

\begin{table}[htb]
\begin{tabular}{|c|c|c|c|c|}
\hline 
$A_{0}$  & $D_{0}$  & $\hbar\Delta_{m}$  & $E_{\omega}$  & $E_{2\omega}$ \tabularnewline
\hline 
$5\cdot10^{5}\frac{m}{s}$  & $7\cdot10^{-4}\frac{m^{2}}{s}$  & $1.5\cdot10^{-2}eV$  & $10^{4}\frac{V}{m}$  & $72\frac{V}{m}$ \tabularnewline
\hline 
\end{tabular}\caption{Values of the parameters used for the plots. }

\label{tab:param} 
\end{table}

We consider field amplitudes of $E_{\omega}=10^{4}\frac{V}{m}$ for
the fundamental and $E_{2\omega}=72\frac{V}{m}$ for the second harmonic,
which are indicative of the largest field intensities allowed within
the perturbative regime. These values depend on the expressions for
the injected carrier density, so we explain how they are obtained
in Sec. \ref{sec:discussion}.

\subsection{Linear polarizations }

The one- and two-photon processes do not depend on the relative orientation
of the fundamental $\boldsymbol{E}\left(\omega\right)=E_{\omega}e^{i\theta_{1}}\hat{\boldsymbol{e}}_{\omega}$
and second harmonic $\boldsymbol{E}\left(2\omega\right)=E_{2\omega}e^{i\theta_{2}}\hat{\boldsymbol{e}}_{2\omega}$
fields, where $E_{\omega}$ and $E_{2\omega}$ are real. Therefore
we show here the results for the injection coefficients $\Lambda_{1}$
and $\Lambda_{2}$, while the results for $\Lambda_{i\left(1\right)}$
and $\Lambda_{i\left(2\right)}$ are displayed for the special cases
of parallel and perpendicular polarizations.

The carrier density injection rate is given by 
\begin{equation}
\begin{array}{rl}
\left\langle \dot{n}_{1}\right\rangle = & \xi_{1}^{xx}\left(\omega\right)E_{\omega}^{2}+\xi_{1}^{xx}\left(2\omega\right)E_{2\omega}^{2},\\
\left\langle \dot{n}_{2}\right\rangle = & \xi_{2}^{xxxx}\left(\omega\right)E_{\omega}^{4},
\end{array}
\end{equation}
and the $\hat{\boldsymbol{z}}$ component of the spin density injection
rate is given by 
\begin{equation}
\begin{array}{rl}
\left\langle \dot{S}_{1}^{z}\right\rangle = & \frac{2\hbar\Delta_{m}}{\omega}\left[\xi_{1}^{xx}\left(\omega\right)E_{\omega}^{2}+\frac{1}{2}\xi_{1}^{xx}\left(2\omega\right)E_{2\omega}^{2}\right],\\
\left\langle \dot{S}_{2}^{z}\right\rangle = & \frac{\hbar\Delta_{m}}{\omega}\xi_{2}^{xxxx}\left(\omega\right)E_{\omega}^{4}.
\end{array}\label{eq:injspinz}
\end{equation}
This result simply corresponds to the net polarization of the injected
carriers.

The charge current injection rate vanishes; $\left\langle \dot{J}_{1}^{a}\right\rangle =\left\langle \dot{J}_{2}^{a}\right\rangle =0$.

The spin current injection rate is 
\begin{equation}
\begin{array}{rl}
\left\langle \dot{J}_{S,1}^{ab}\right\rangle = & \underset{n=1,2}{\sum}\left(\hat{\boldsymbol{z}}\times\hat{\boldsymbol{e}}_{n\omega}\right)\cdot\hat{\boldsymbol{a}}\left(\hat{\boldsymbol{e}}_{n\omega}\cdot\hat{\boldsymbol{b}}\right)\mu_{1}^{yxxx}\left(n\omega\right)E_{n\omega}^{2}+\\
 & +\underset{n=1,2}{\sum}\hat{\boldsymbol{e}}_{n\omega}\cdot\hat{\boldsymbol{a}}\left(\hat{\boldsymbol{z}}\times\hat{\boldsymbol{e}}_{n\omega}\right)\cdot\hat{\boldsymbol{b}}\mu_{1}^{xyxx}\left(n\omega\right)E_{n\omega}^{2},\\
\left\langle \dot{J}_{S,2}^{ab}\right\rangle = & \left(\hat{\boldsymbol{z}}\times\hat{\boldsymbol{e}}_{n\omega}\right)\cdot\hat{\boldsymbol{a}}\left(\hat{\boldsymbol{e}}_{n\omega}\cdot\hat{\boldsymbol{b}}\right)\mu_{2}^{yxxxxx}\left(\omega\right)E_{\omega}^{4}+\\
 & +\hat{\boldsymbol{e}}_{\omega}\cdot\hat{\boldsymbol{a}}\left(\hat{\boldsymbol{z}}\times\hat{\boldsymbol{e}}_{\omega}\right)\cdot\hat{\boldsymbol{b}}\mu_{2}^{xyxxxx}\left(\omega\right)E_{\omega}^{4},
\end{array}\label{eq:linearSpinCurr1}
\end{equation}
the first term in each equation gives a spin current independent of
the applied field polarization, and is due the helical spin structure.

\subsubsection*{Parallel orientations }

Only the interference processes depend on the relative orientation
of the $\boldsymbol{E}\left(\omega\right)$ and $\boldsymbol{E}\left(2\omega\right)$.
Here the fields are $\boldsymbol{E}\left(\omega\right)=E_{\omega}e^{i\theta_{1}}\hat{\boldsymbol{e}}_{\omega}$
and $\boldsymbol{E}\left(2\omega\right)=E_{2\omega}e^{i\theta_{2}}\hat{\boldsymbol{e}}_{\omega}$.
The relative phase parameter is $\Delta\theta=\theta_{2}-2\theta_{1}$.

The charge current injection rate is given by 
\begin{equation}
\begin{array}{rl}
\left\langle \dot{\boldsymbol{J}}_{i}\right\rangle = & -2\hat{\boldsymbol{e}}_{\omega}{\rm Im}\left[\eta_{i\left(1\right)}^{xxxx}\left(\omega\right)+\eta_{i\left(2\right)}^{xxxx}\left(\omega\right)\right]\\
 & \cdot\sin\left(\Delta\theta\right)E_{\omega}^{2}E_{2\omega}.
\end{array}
\end{equation}
Due to Eq. \eqref{eq:relations}, the in plane spin density and the
spin $\hat{\boldsymbol{z}}$ current injection rates are given in
terms of $\left\langle \dot{\boldsymbol{J}}_{i}\left(\omega\right)\right\rangle $
by 
\begin{equation}
\begin{array}{rl}
\left\langle \dot{\boldsymbol{S}}_{i}\right\rangle = & \frac{\hbar}{2A_{0}e}\hat{\boldsymbol{z}}\times\left\langle \dot{\boldsymbol{J}}_{i}\right\rangle ,\\
\left\langle \dot{\boldsymbol{J}}_{S,i}^{z}\right\rangle = & \frac{\hbar D_{0}\Delta_{m}}{A_{0}^{2}e}\left\langle \dot{\boldsymbol{J}}_{i}\right\rangle ,
\end{array}\label{eq:injecrel}
\end{equation}
the spin current merely corresponds to the magnetization of the carriers
of the charge current.

The direction of the polarization vector provides control of the angle
of the injected vectorial quantities, while the relative phase parameter
of the light beams can control only their magnitude and orientation.

\subsubsection*{Perpendicular orientations }

Here we have $\boldsymbol{E}\left(\omega\right)=E_{\omega}e^{i\theta_{1}}\hat{\boldsymbol{e}}_{\omega}$
and $\boldsymbol{E}\left(2\omega\right)=E_{2\omega}e^{i\theta_{2}}\hat{\boldsymbol{e}}_{2\omega}$
with $\hat{\boldsymbol{e}}_{2\omega}=\hat{\boldsymbol{z}}\times\hat{\boldsymbol{e}}_{\omega}$.
The relative phase parameter is again $\Delta\theta=\theta_{2}-2\theta_{1}$.

The charge current injection rate is given by 
\begin{equation}
\begin{array}{rl}
\left\langle \dot{J}_{i}^{a}\right\rangle = & -2\hat{\boldsymbol{e}}_{2\omega}\cdot\hat{\boldsymbol{a}}{\rm Im}\left[\eta_{i\left(1\right)}^{yxxy}\left(\omega\right)+\eta_{i\left(2\right)}^{yxxy}\left(\omega\right)\right]\\
 & \cdot\sin\left(\Delta\theta\right)E_{\omega}^{2}E_{2\omega}\\
 & +2\hat{\boldsymbol{e}}_{\omega}\cdot\hat{\boldsymbol{a}}{\rm Re}\left[\eta_{i\left(1\right)}^{xxxy}\left(\omega\right)+\eta_{i\left(2\right)}^{xxxy}\left(\omega\right)\right]\\
 & \cdot\cos\left(\Delta\theta\right)E_{\omega}^{2}E_{2\omega},
\end{array}
\end{equation}
and the spin density and current follow Eq. \eqref{eq:injecrel}.

From Eqs. \eqref{eq:currcoeff1} and \eqref{eq:currcoeff2} in Appendix
\ref{app:coeffs} we can identify two different contributions to the
injection: one that is related to the Berry curvature, and thus depends
on $\Delta_{m}$, and another that is independent of $\Delta_{m}$. 

Again the direction of the polarization vector provides control of
the angle of the injected vectorial quantities. The relative phase
can still control their magnitude and orientation, but it can also
switch between the two regimes: the first where the photoinjection
stems from the Berry curvature, and the second where it does not.

\begin{figure}[htbp]
\begin{tabular}{rr}
\includegraphics[width=0.49\columnwidth]{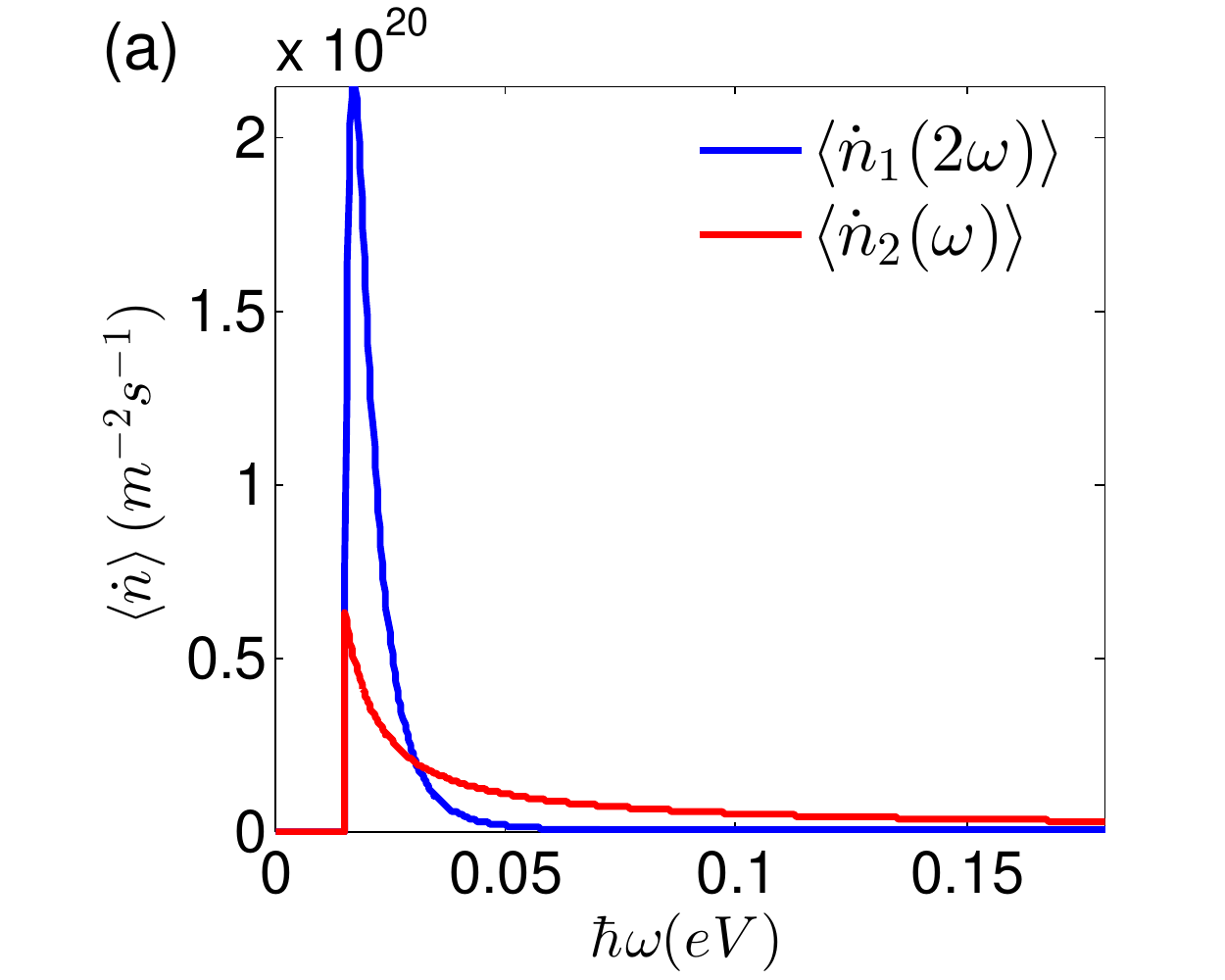}  & \includegraphics[width=0.49\columnwidth]{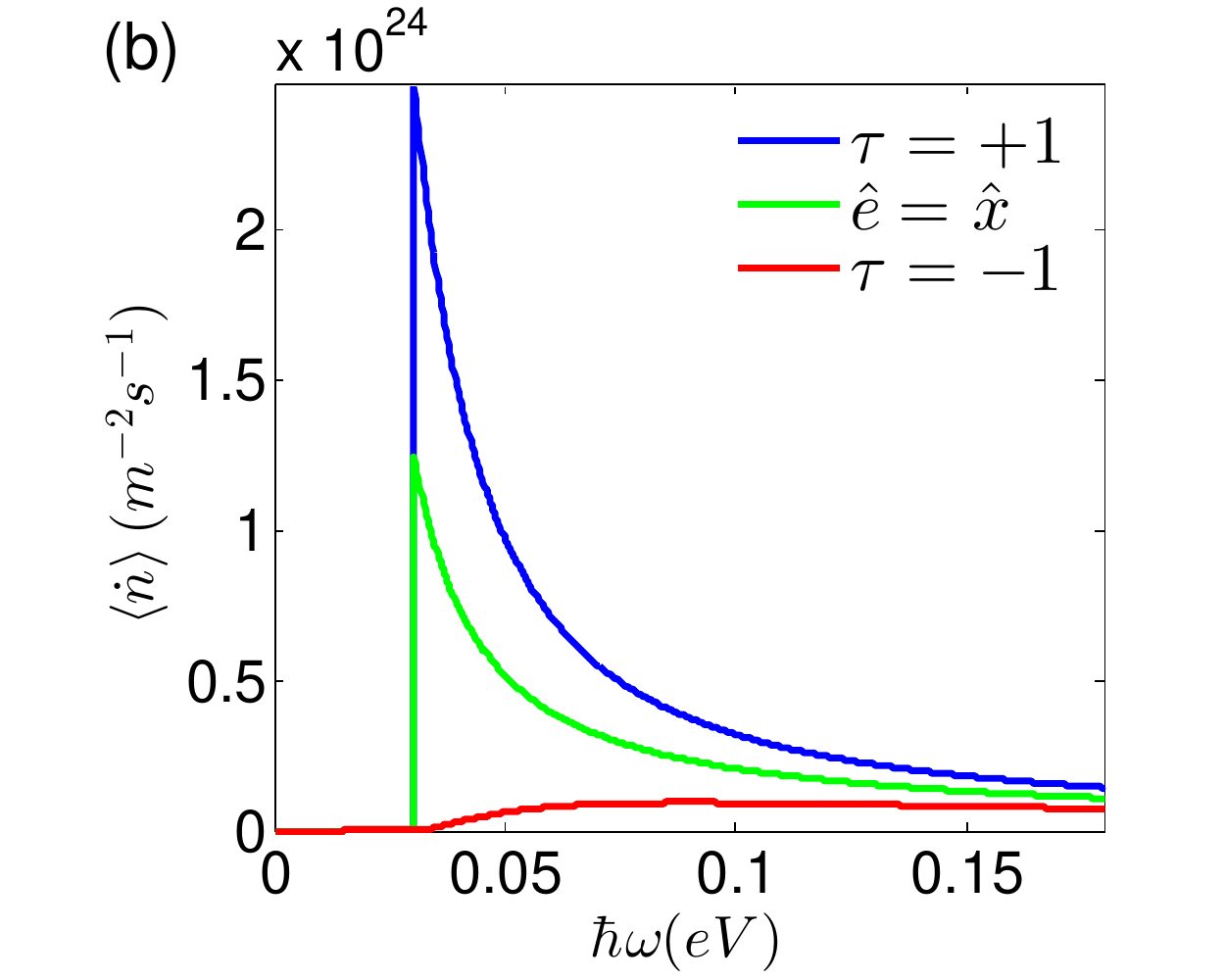} \tabularnewline
\end{tabular}\caption[]{(Color online) (a) Carrier density injection rates from one- and
two-photon absorption processes at total energy $2\hbar\omega$. (b)
Carrier density injection rates for linear($\hat{\boldsymbol{e}}_{\omega}=\hat{\boldsymbol{x}}$)
and circular($\tau=\pm1$) polarizations of the incident fields. }

\label{fig:carrier} 
\end{figure}

\begin{figure}[htbp]
\begin{tabular}{rr}
\includegraphics[width=0.49\columnwidth]{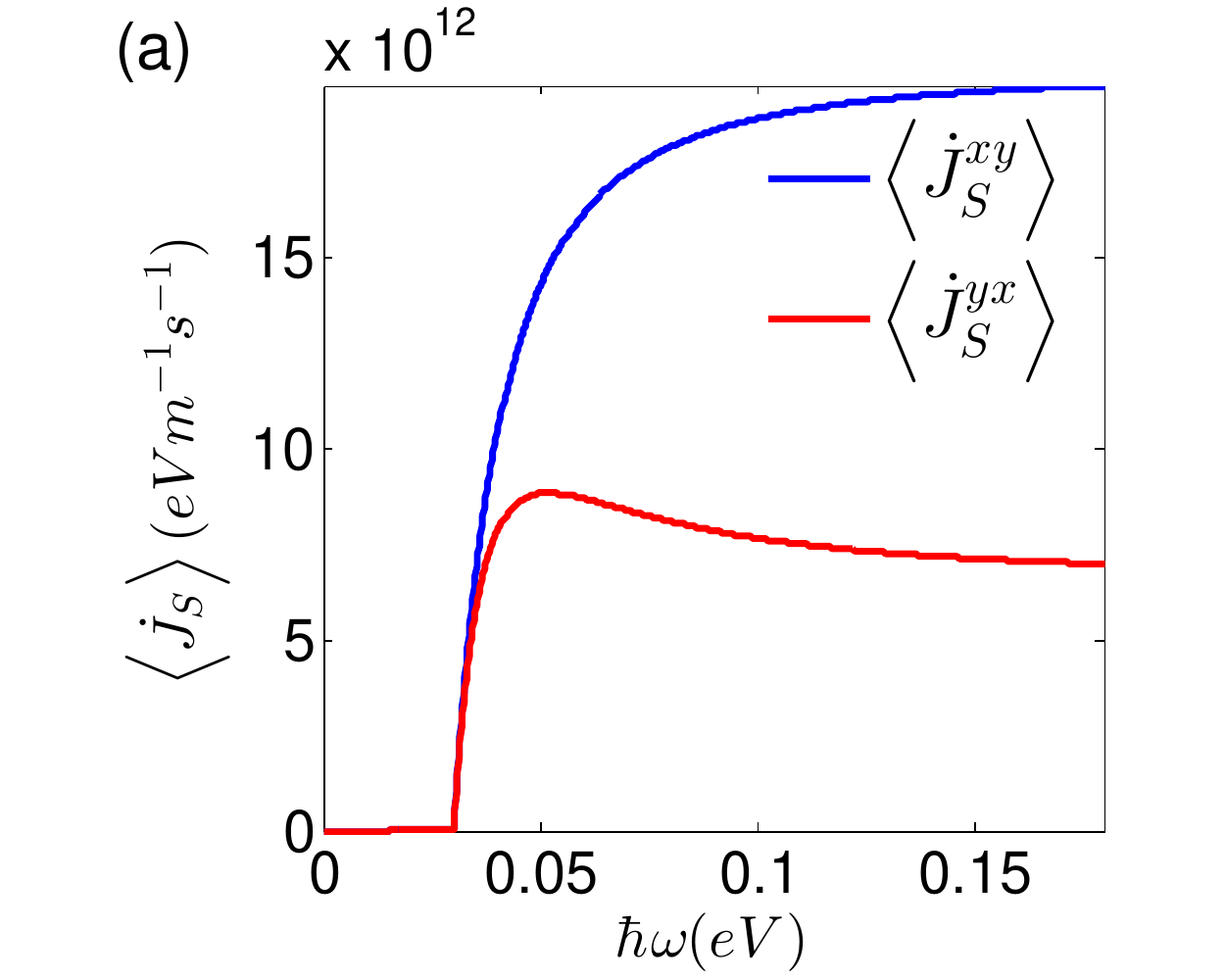}  & \includegraphics[width=0.49\columnwidth]{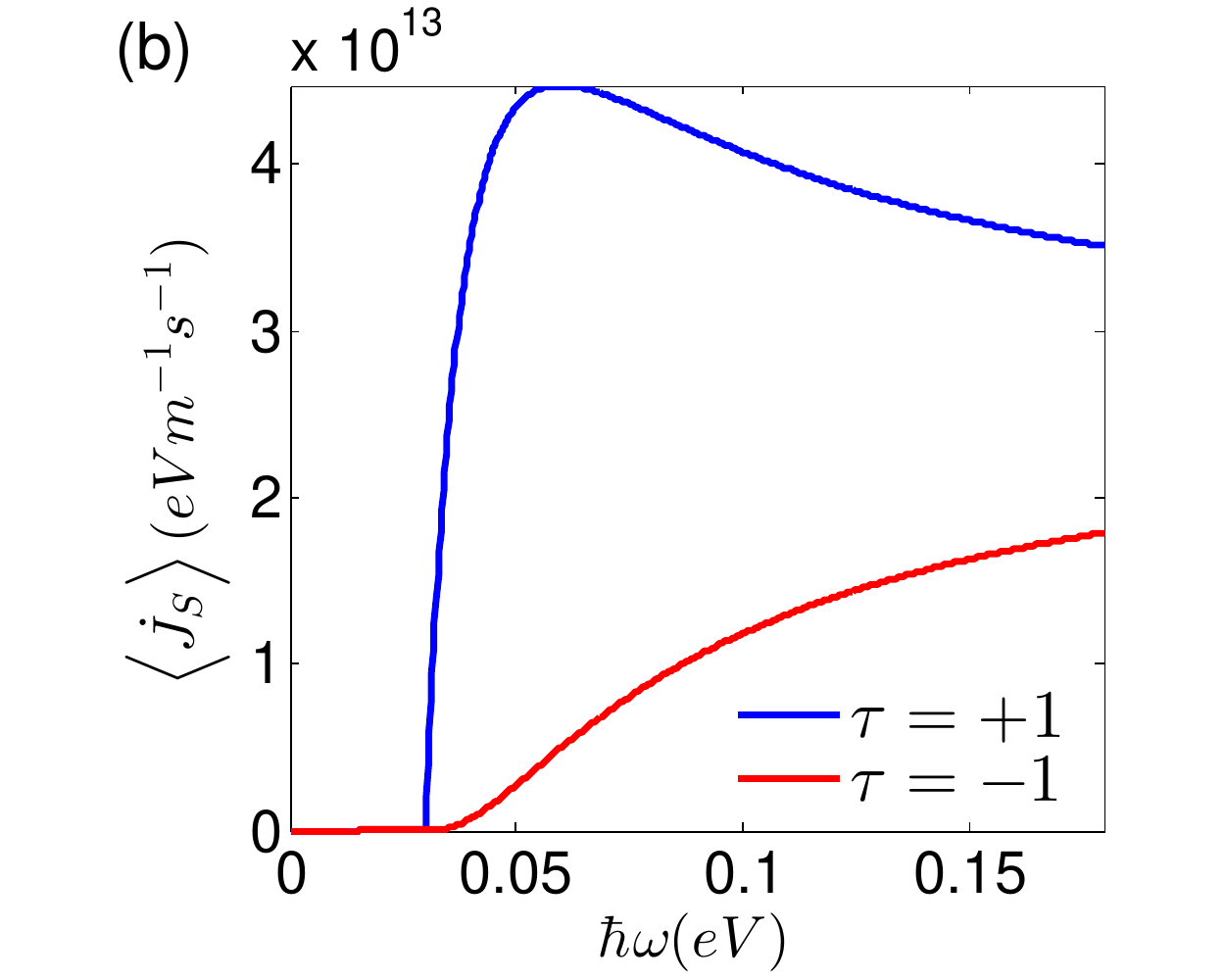} \tabularnewline
\end{tabular}\caption[]{(Color online) Planar spin current density injection rates for (a)
linear polarizations along the $\hat{\boldsymbol{x}}$ direction and
(b) circular polarizations. }

\label{fig:spincurr} 
\end{figure}

\begin{figure*}[bthp]
\begin{tabular}{rrr}
\includegraphics[width=0.33\textwidth]{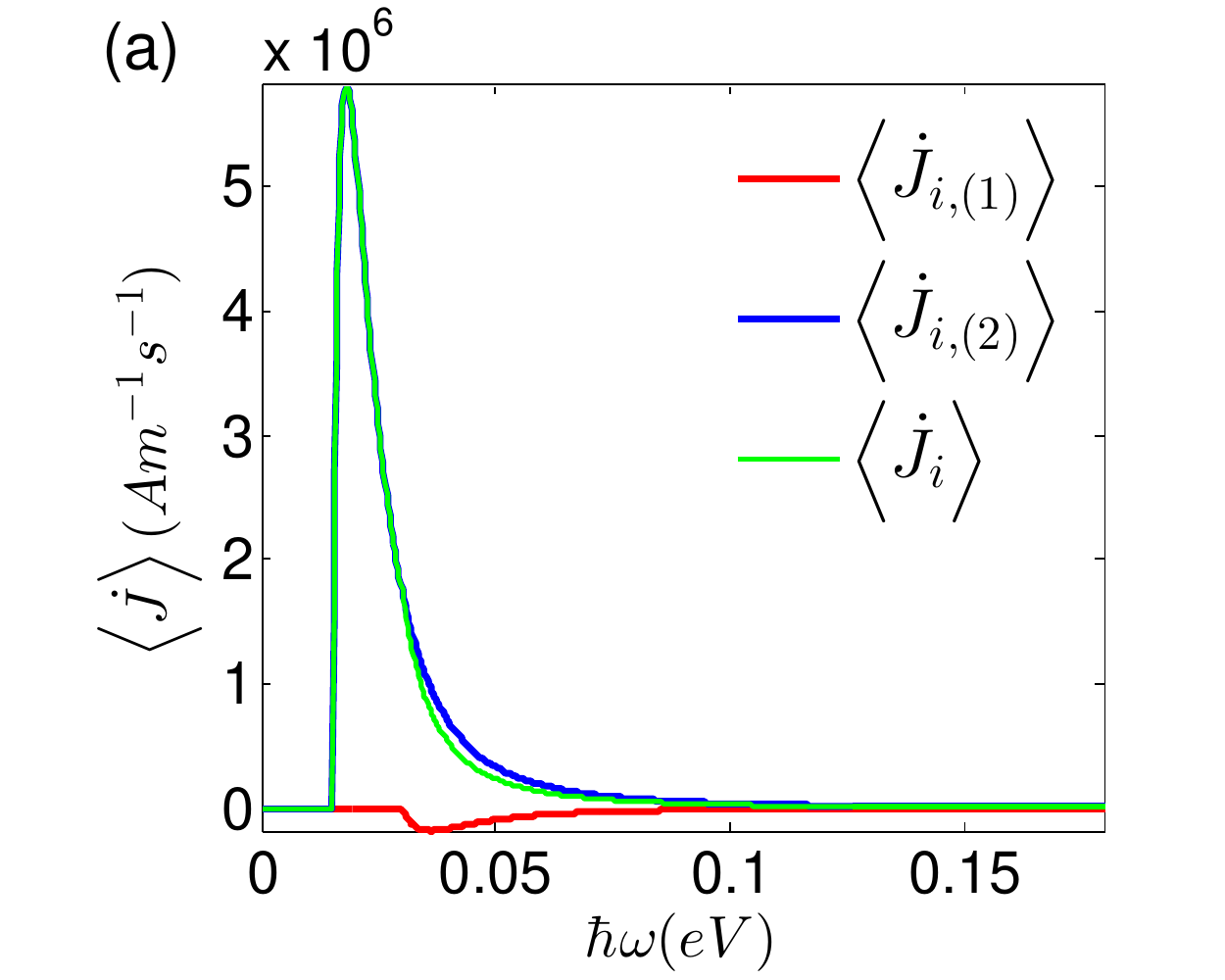}  & \includegraphics[width=0.33\textwidth]{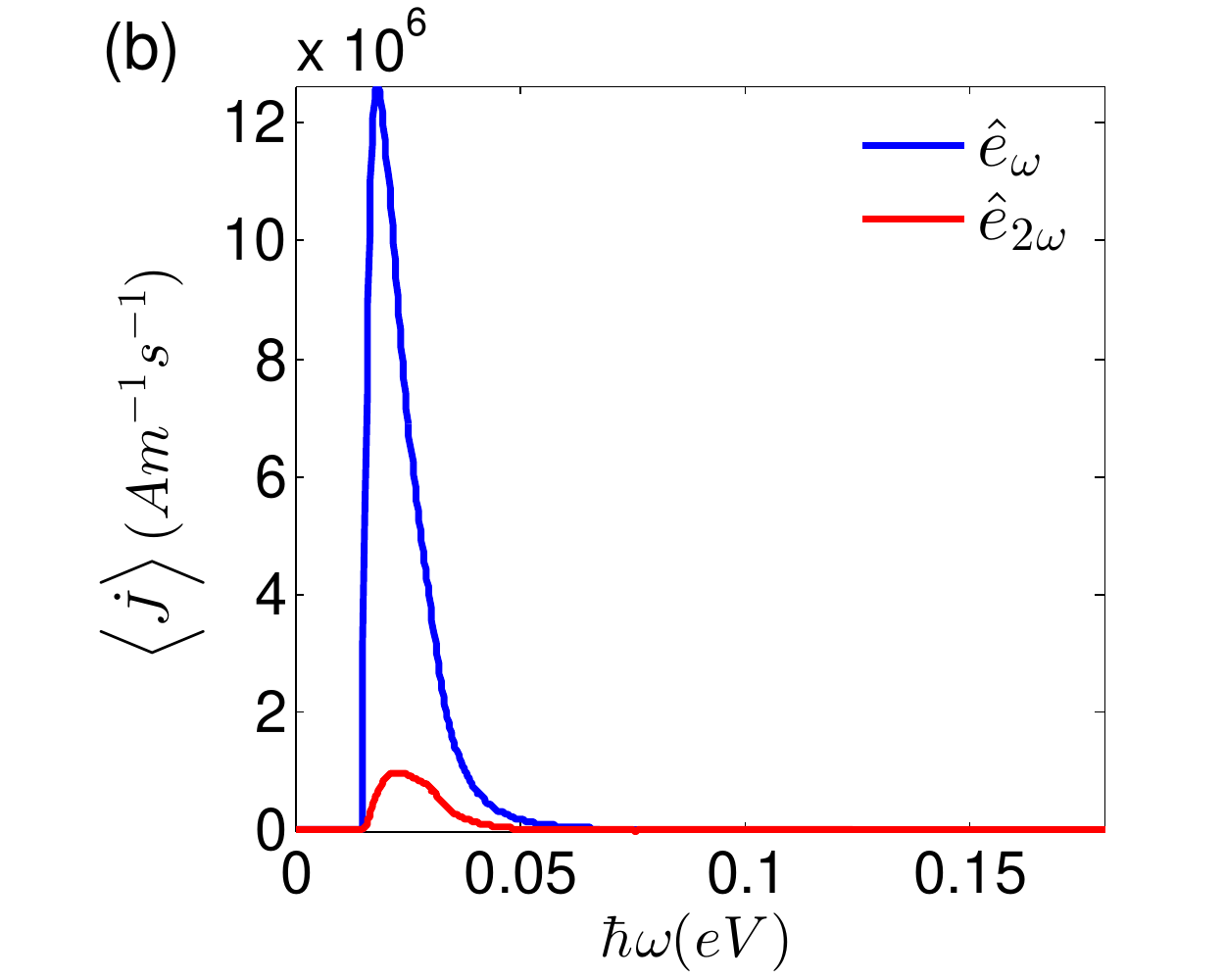}  & \includegraphics[width=0.33\textwidth]{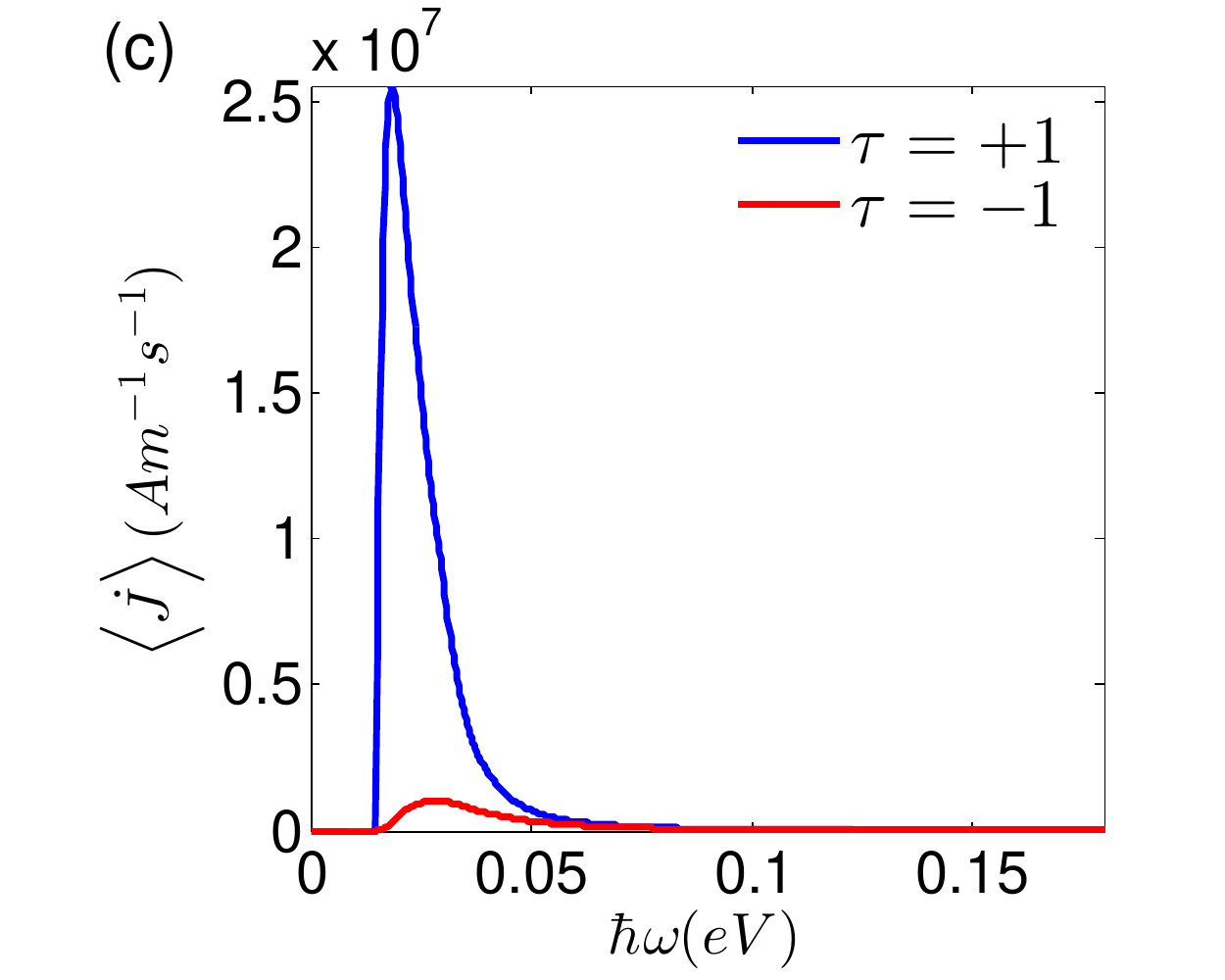} \tabularnewline
\end{tabular}\caption[]{(Color online) Current density injection rates for (a) linear polarizations
with parallel orientations (b) linear polarizations with perpendicular
orientations, showing the components of the current along the $\hat{\boldsymbol{e}}_{\omega}$
and $\hat{\boldsymbol{e}}_{2\omega}$ directions, and (c) circular
polarizations. }

\label{fig:current} 
\end{figure*}

\subsection{Circular polarizations }

For circular polarizations $\boldsymbol{E}\left(\omega\right)=E_{\omega}e^{i\theta_{1}}\hat{\boldsymbol{p}}_{\tau_{1}}$
and $\boldsymbol{E}\left(2\omega\right)=E_{2\omega}e^{i\theta_{2}}\hat{\boldsymbol{p}}_{\tau_{2}}$
where $\tau_{1},\tau_{2}=\pm1$ and $\hat{\boldsymbol{p}}_{\pm}=\left(\hat{\boldsymbol{x}}\pm i\hat{\boldsymbol{y}}\right)/\sqrt{2}$,
so $\hat{\boldsymbol{p}}_{\tau}\cdot\hat{\boldsymbol{p}}_{\tau}=0$
and $\hat{\boldsymbol{p}}_{+}\cdot\hat{\boldsymbol{p}}_{-}=1$ as
well as $\hat{\boldsymbol{p}}_{-}\times\hat{\boldsymbol{p}}_{+}=i\hat{\boldsymbol{z}}$.
The relative phase parameter is still $\Delta\theta=\theta_{2}-2\theta_{1}$.
Again the one and two photons processes do not depend on the relative
helicity of the $\boldsymbol{E}\left(\omega\right)$ and $\boldsymbol{E}\left(2\omega\right)$
fields and are presented first.

The carrier density injection rate is now given by 
\begin{equation}
\begin{array}{rl}
\left\langle \dot{n}_{1}\right\rangle = & \xi_{1}^{-+}\left(\omega\right)E_{\omega}^{2}+\xi_{1}^{-+}\left(2\omega\right)E_{2\omega}^{2},\\
\left\langle \dot{n}_{2}\right\rangle = & \xi_{2}^{--++}\left(\omega\right)E_{\omega}^{4}.
\end{array}
\end{equation}
The spin density injection is still given by Eq. \eqref{eq:injspinz},
and for the spin current we have 
\begin{equation}
\begin{array}{rl}
\left\langle \dot{J}_{S,1}^{ab}\right\rangle = & 2i\left(\hat{\boldsymbol{a}}\times\hat{\boldsymbol{b}}\right)\cdot\hat{\boldsymbol{z}}\underset{n=1,2}{\sum}\mu_{1}^{-+-+}\left(n\omega\right)E_{n\omega}^{2},\\
\left\langle \dot{J}_{S,2}^{ab}\right\rangle = & 2i\left(\hat{\boldsymbol{a}}\times\hat{\boldsymbol{b}}\right)\cdot\hat{\boldsymbol{z}}\mu_{2}^{-+--++}\left(\omega\right)E_{\omega}^{4}.
\end{array}
\end{equation}
Circularly polarized light does not break rotational symmetry, therefore
the second term of Eq. \eqref{eq:linearSpinCurr1} is not present.

\subsubsection*{Equal helicities }

The interference processes depend on the relative helicity of the
two fields. We first consider the fields with the same helicity, $\boldsymbol{E}\left(\omega\right)=E_{\omega}e^{i\theta_{1}}\hat{\boldsymbol{p}}_{\tau}$
and $\boldsymbol{E}\left(2\omega\right)=E_{2\omega}e^{i\theta_{2}}\hat{\boldsymbol{p}}_{\tau}$.

The charge current injection rate is given by 
\begin{equation}
\begin{array}{rl}
\left\langle \dot{J}_{i}^{a}\right\rangle = & 2\left[a^{y}\cos\left(\Delta\theta\right)+a^{x}\sin\left(\Delta\theta\right)\right]\\
 & \cdot i\left[\eta_{i\left(1\right)}^{+--+}\left(\omega\right)+\eta_{i\left(2\right)}^{+--+}\left(\omega\right)\right]E_{\omega}^{2}E_{2\omega}.
\end{array}
\end{equation}
The spin density and current follows Eq. \eqref{eq:injecrel}.

The relative phase displacement between the two light beams can now
control the direction of the injected quantities.

Especially for frequencies near the gap, the injection rates for different
helicities $\tau$ depend strongly on the chirality of the electronic
states, identified by $\Delta_{m}/\left|\Delta_{m}\right|$. The helicity
of the incident light has no effect for vanishing $\Delta_{m}$.

\subsubsection*{Opposite helicities }

Here we have $\boldsymbol{E}\left(\omega\right)=E_{\omega}e^{i\theta_{1}}\hat{\boldsymbol{p}}_{\tau}$
and $\boldsymbol{E}\left(2\omega\right)=E_{2\omega}e^{i\theta_{2}}\hat{\boldsymbol{p}}_{-\tau}$.
The injection rates from interference all vanish for the four operators
of interest.

\section{Discussion }

\label{sec:discussion}

In order to determine the validity our calculations for the optical
injection rates, we have to consider the fraction of the injected
carrier population relative to the total number of states in the range
of energies covered by the laser pulse. The duration of the pulse $\cal{T} $
sets the frequency broadening of the laser $\Delta \omega = \frac{2\pi}{\cal T}$, which in turn - via the
dispersion relation, which we assume $ E_{\boldsymbol{k}} = \hbar A_0 k$ here for simplicity  - determines the area of the Brillouin zone that
can be populated by carriers $a=2\pi k \Delta k$, where $k=\frac{\omega}{A_0}$ and $\Delta k = \frac{\Delta \omega}{A_0}$.
The number of states available in this area of the Brillouin zone is $a / a_1$, where $a_1 = \frac{\left(2\pi\right)^2}{L^2}$ is the area occupied by one state. 
The maximum amplitudes of the laser fields are restricted by  the condition that
the number of injected carriers with additional energy  $2\hbar \omega$ is at most $5\%$ of the total number 
of carrier states in the allowed energy range
\begin{equation}
\left( \xi_{1}^{xx}\left(2\omega\right)E_{2\omega}^{2}+\xi_{2}^{xxxx}\left(\omega\right)E_{\omega}^{4} \right){\cal T} L^2  < 0.05 \frac{a}{a_1}.
\label{eq:condition}
\end{equation} 
We then estimate the amplitudes by imposing the additional condition
$\xi_{1}^{xx}\left(2\omega\right)E_{2\omega}^{2}=\xi_{2}^{xxxx}\left(\omega\right)E_{\omega}^{4}$,
which gives optimal interference between the absorption processes \cite{rioux12}. 
Finally, the field amplitudes are limited by  
\begin{equation}
 \left(eE_{2\omega}\right)^{2}=\frac{4 A_{0}^{2}\left(eE_{\omega}\right)^{4}}{\hbar^{2}\omega^{4}}<\frac{1.6\hbar^{2}\omega^{2}}{A_{0}^{2}{\cal T}^{2}}. 
 \label{eq:maxamp}
\end{equation}
For pulses lasting $1ns$ with a frequency of $30meV$,
the field amplitudes found are $E_{\omega}=10^{4}\frac{V}{m}$ for
the fundamental and $E_{2\omega}=72\frac{V}{m}$ for the second harmonic,
which correspond to laser intensities of $9.9\frac{W}{cm^{2}}$ and
$0.65\frac{mW}{cm^{2}}$, respectively. 
We use these values for all $\hbar \omega$ in Figs. \ref{fig:carrier},\ref{fig:spincurr}, and \ref{fig:current}, although for $\hbar \omega < 30 meV$ smaller amplitudes would be required to guarantee Eq. \eqref{eq:condition}, and could be found by using Eq. \eqref{eq:maxamp}. 

In the absence of the $\sigma_{z}$ mass term, the carrier and charge
current injection rates are very similar to the ones found for graphene
\cite{rioux11}, except for the adjustments due to having only one
Dirac cone and a smaller Fermi velocity. However, even in this case
there is also injection of the transverse spin following the same
form of the injected current, a signature characteristic of topological
insulators. The magnitude of these injected quantities are also of
the same order of the values for graphene, which has already been
measured \cite{norris10}.

Another distinctive trait shared with graphene is the relatively low
average velocity of the injected carriers when compared to semiconductors.
This is due to the one photon absorption at the fundamental frequency,
forbidden in semiconductors because of the bandgap. This gives rise
to the extra interference process with total energy $\hbar\omega$,
which usually partially cancels the injected current stemming from
the interference process with total energy $2\hbar\omega$.

Several particular features are found in the presence of the Berry
phase inducing $\Delta_{m}\sigma_{z}$ term, especially for circular
polarizations of the optical fields, when an interesting interplay
between the helicity $\tau$ of the incident fields and the chirality
$\Delta_{m}/\left|\Delta_{m}\right|$ of the Dirac cone can greatly
suppress or enhance optical injection. In order to observe these features
a combination of high magnetic field and low temperature is necessary.
Because the Zeeman coupling $\hbar\Delta_{m}=\frac{\mu_{B}g}{2}B$
needs to be above temperature $k_{B}T$. We estimate that $77K$ and
$6T$ should be enough for $Bi_{2}Te_{3}$. For a pronounced effect,
the photon energy should not be much larger than the Zeeman gap. Reasonable
photon energies for the fundamental field would not be much larger
than $\hbar\omega=30meV$, which can be achieved with quantum cascade
lasers.

When lasers of similar intensity are considered, the magnitude of
the injected currents obtained from coherent control seem to be considerably
larger than the values found by other approaches, like applying an
in-plane magnetic field or oblique incidence. Therefore it can play
a crucial role in the quest for harnessing the exotic properties of
topological insulators for spintronic applications.

\acknowledgments

We thank Julien Rioux and Jin Luo Cheng for helpful discussions. This
work was supported by the Natural Sciences and Engineering Research
Council of Canada (NSERC).

\appendix

\section{General expressions for the optical injection coefficients }

\label{app:formula}

In order to evaluate the coefficients in Eq. \eqref{eq:coeffinal},
the following integrals are helpful 
\begin{equation}
\begin{array}{rl}
\varphi^{ab}= & \int\frac{d\theta}{2\pi}\frac{k^{a}k^{b}}{k^{2}}=\frac{\hat{\boldsymbol{a}}\cdot\hat{\boldsymbol{b}}}{2},\\
\varphi^{abcd}= & \int\frac{d\theta}{2\pi}\frac{k^{a}k^{b}k^{c}k^{d}}{k^{4}}=\frac{\hat{\boldsymbol{a}}\cdot\hat{\boldsymbol{b}}\left(\hat{\boldsymbol{c}}\cdot\hat{\boldsymbol{d}}\right)+\hat{\boldsymbol{a}}\cdot\hat{\boldsymbol{c}}\left(\hat{\boldsymbol{b}}\cdot\hat{\boldsymbol{d}}\right)+\hat{\boldsymbol{a}}\cdot\hat{\boldsymbol{d}}\left(\hat{\boldsymbol{b}}\cdot\hat{\boldsymbol{c}}\right)}{8},\\
\varphi^{abcdef}= & \int\frac{d\theta}{2\pi}\frac{k^{a}k^{b}k^{c}k^{d}k^{e}k^{f}}{k^{6}}=\underset{pairings}{\sum}\frac{\hat{\boldsymbol{a}}\cdot\hat{\boldsymbol{b}}\left(\hat{\boldsymbol{c}}\cdot\hat{\boldsymbol{d}}\right)\hat{\boldsymbol{e}}\cdot\hat{\boldsymbol{f}}}{48},
\end{array}
\end{equation}
and 
\begin{equation}
\begin{array}{rl}
\int\frac{d\theta}{2\pi}v_{cv}^{c}v_{vc}^{b}= & \frac{A_{0}^{2}\left(d_{\boldsymbol{k}}^{2}+\Delta_{m}^{2}\right)\hat{\boldsymbol{c}}\cdot\hat{\boldsymbol{b}}}{2d_{\boldsymbol{k}}^{2}}+i\frac{A_{0}^{2}\Delta_{m}\hat{\boldsymbol{z}}\cdot\left(\hat{\boldsymbol{c}}\times\hat{\boldsymbol{b}}\right)}{d_{\boldsymbol{k}}},\\
\int\frac{d\theta}{2\pi}k^{a}v_{cv}^{c}v_{vc}^{b}= & 0,
\end{array}
\end{equation}
also 
\begin{equation}
\begin{array}{rl}
\int\frac{d\theta}{2\pi}k^{d}k^{b}v_{cv}^{c}v_{vc}^{e}= & \frac{\left(d_{\boldsymbol{k}}^{2}-\Delta_{m}^{2}\right)\left[d_{\boldsymbol{k}}^{2}\hat{\boldsymbol{c}}\cdot\hat{\boldsymbol{e}}\left(\hat{\boldsymbol{d}}\cdot\hat{\boldsymbol{b}}\right)-2\left(d_{\boldsymbol{k}}^{2}-\Delta_{m}^{2}\right)\varphi^{bcde}\right]}{2d_{\boldsymbol{k}}^{2}}\\
 & +i\frac{\left(d_{\boldsymbol{k}}^{2}-\Delta_{m}^{2}\right)\Delta_{m}\hat{\boldsymbol{d}}\cdot\hat{\boldsymbol{b}}\left(\hat{\boldsymbol{c}}\times\hat{\boldsymbol{e}}\right)\cdot\hat{\boldsymbol{z}}}{2d_{\boldsymbol{k}}},
\end{array}
\end{equation}
and 
\begin{equation}
\begin{array}{rl}
\int\frac{d\theta}{2\pi}k^{c}k^{d}k^{e}v_{cv}^{a}v_{vc}^{b}= & 0,\\
\int\frac{d\theta}{2\pi}k^{c}k^{d}k^{e}k^{f}v_{cv}^{a}v_{vc}^{b}= & \frac{\left(d_{\boldsymbol{k}}^{2}-\Delta_{m}^{2}\right)^{2}\left[d_{\boldsymbol{k}}^{2}\hat{\boldsymbol{a}}\cdot\hat{\boldsymbol{b}}\varphi^{cdef}-\left(d_{\boldsymbol{k}}^{2}-\Delta_{m}^{2}\right)\varphi^{abcdef}\right]}{A_{0}^{2}d_{\boldsymbol{k}}^{2}}\\
 & +i\frac{\left(d_{\boldsymbol{k}}^{2}-\Delta_{m}^{2}\right)^{2}\Delta_{m}\left(\hat{\boldsymbol{a}}\times\hat{\boldsymbol{b}}\right)\cdot\hat{\boldsymbol{z}}\varphi^{cdef}}{A_{0}^{2}d_{\boldsymbol{k}}}.
\end{array}
\end{equation}
The above equations combined with Eq. \eqref{eq:ticomps}, Eq. \eqref{eq:tivelvel},
Eq. \eqref{eq:tigamma12} and Eq. \eqref{eq:tigammaint} gives the
following results.

\begin{widetext}

\subsubsection*{One and two photon absorption }

The carrier density coefficients are 
\begin{equation}
\begin{array}{rl}
\xi_{1}^{bc}\left(\omega\right)= & \frac{\Theta\left(\omega-2\Delta_{m}\right)e^{2}}{2\hbar^{2}\omega}\left[\frac{\hat{\boldsymbol{b}}\cdot\hat{\boldsymbol{c}}}{4}\left(1+\frac{4\Delta_{m}^{2}}{\omega^{2}}\right)-i\frac{\Delta_{m}\hat{\boldsymbol{z}}\cdot\left(\hat{\boldsymbol{b}}\times\hat{\boldsymbol{c}}\right)}{\omega}\right],\\
\xi_{2}^{bcde}\left(\omega\right)= & \frac{\Theta\left(\omega-\Delta_{m}\right)e^{4}A_{0}^{2}}{\hbar^{4}\omega^{5}}\left(1-\frac{\Delta_{m}^{2}}{\omega^{2}}\right)\left(\left[\frac{\left(\hat{\boldsymbol{b}}\cdot\hat{\boldsymbol{d}}\right)\hat{\boldsymbol{c}}\cdot\hat{\boldsymbol{e}}+\left(\hat{\boldsymbol{b}}\cdot\hat{\boldsymbol{e}}\right)\hat{\boldsymbol{c}}\cdot\hat{\boldsymbol{d}}}{2}-2\varphi^{bcde}\left(1-\frac{\Delta_{m}^{2}}{\omega^{2}}\right)\right]+i\frac{\Delta_{m}}{\omega}\left[\frac{\hat{\boldsymbol{c}}\cdot\hat{\boldsymbol{e}}\left(\hat{\boldsymbol{d}}\times\hat{\boldsymbol{b}}\right)\cdot\hat{\boldsymbol{z}}+\hat{\boldsymbol{b}}\cdot\hat{\boldsymbol{e}}\left(\hat{\boldsymbol{d}}\times\hat{\boldsymbol{c}}\right)\cdot\hat{\boldsymbol{z}}+\hat{\boldsymbol{c}}\cdot\hat{\boldsymbol{d}}\left(\hat{\boldsymbol{e}}\times\hat{\boldsymbol{b}}\right)\cdot\hat{\boldsymbol{z}}+\hat{\boldsymbol{d}}\cdot\hat{\boldsymbol{b}}\left(\hat{\boldsymbol{e}}\times\hat{\boldsymbol{c}}\right)\cdot\hat{\boldsymbol{z}}}{4}\right]\right).
\end{array}
\end{equation}

The charge current coefficients vanish, $\eta_{1}^{abc}\left(\omega\right)=\eta_{2}^{abcde}\left(\omega\right)=0$.

The spin density coefficients can be written in terms of the carrier
density ones as 
\begin{equation}
\begin{array}{rl}
\zeta_{1}^{abc}\left(\omega\right)= & \frac{2\hbar\Delta_{m}\left(\hat{\boldsymbol{z}}\cdot\hat{\boldsymbol{a}}\right)}{\omega}\xi_{1}^{bc}\left(\omega\right),\\
\zeta_{2}^{abcde}\left(\omega\right)= & \frac{\hbar\Delta_{m}\left(\hat{\boldsymbol{z}}\cdot\hat{\boldsymbol{a}}\right)}{\omega}\xi_{2}^{bcde}\left(\omega\right),
\end{array}
\end{equation}
which is a consequence of Eq. \eqref{eq:relations}.

And the spin current coefficients are 
\begin{equation}
\begin{array}{rl}
\mu_{1}^{abcd}\left(\omega\right)= & \frac{\Theta\left(\omega-2\Delta_{m}\right)e^{2}D_{0}}{4A_{0}\hbar}\left(1-\frac{4\Delta_{m}^{2}}{\omega^{2}}\right)\left[\frac{\left(\hat{\boldsymbol{z}}\times\hat{\boldsymbol{a}}\right)\cdot\hat{\boldsymbol{b}}\left(\hat{\boldsymbol{c}}\cdot\hat{\boldsymbol{d}}\right)}{2}-\varphi^{\left(z\times a\right)bcd}\left(1-\frac{4\Delta_{m}^{2}}{\omega^{2}}\right)+i\frac{\Delta_{m}\left(\hat{\boldsymbol{z}}\times\hat{\boldsymbol{a}}\right)\cdot\hat{\boldsymbol{b}}\left(\hat{\boldsymbol{d}}\times\hat{\boldsymbol{c}}\right)\cdot\hat{\boldsymbol{z}}}{\omega}\right],\\
\mu_{2}^{abcdef}\left(\omega\right)= & \frac{4\Theta\left(\omega-\Delta_{m}\right)e^{4}D_{0}A_{0}}{\hbar^{3}\omega^{4}}\left(1-\frac{\Delta_{m}^{2}}{\omega^{2}}\right)^{2}\left[\frac{\varphi^{\left(z\times a\right)bdf}\hat{\boldsymbol{c}}\cdot\hat{\boldsymbol{e}}+\varphi^{\left(z\times a\right)bcf}\hat{\boldsymbol{d}}\cdot\hat{\boldsymbol{e}}+\varphi^{\left(z\times a\right)bde}\hat{\boldsymbol{c}}\cdot\hat{\boldsymbol{f}}+\varphi^{\left(z\times a\right)bce}\hat{\boldsymbol{d}}\cdot\hat{\boldsymbol{f}}}{4}-\varphi^{\left(z\times a\right)bcdef}\left(1-\frac{\Delta_{m}^{2}}{\omega^{2}}\right)\right]\\
 & +i\frac{4\Theta\left(\omega-\Delta_{m}\right)e^{4}D_{0}A_{0}}{\hbar^{3}\omega^{4}}\frac{\Delta_{m}}{\omega}\left(1-\frac{\Delta_{m}^{2}}{\omega^{2}}\right)^{2}\left[\frac{\varphi^{\left(z\times a\right)bdf}\left(\hat{\boldsymbol{e}}\times\hat{\boldsymbol{c}}\right)\cdot\hat{\boldsymbol{z}}+\varphi^{\left(z\times a\right)bcf}\left(\hat{\boldsymbol{e}}\times\hat{\boldsymbol{d}}\right)\cdot\hat{\boldsymbol{z}}+\varphi^{\left(z\times a\right)bde}\left(\hat{\boldsymbol{f}}\times\hat{\boldsymbol{c}}\right)\cdot\hat{\boldsymbol{z}}+\varphi^{\left(z\times a\right)bce}\left(\hat{\boldsymbol{f}}\times\hat{\boldsymbol{d}}\right)\cdot\hat{\boldsymbol{z}}}{4}\right],
\end{array}
\end{equation}
giving spin currents with only in-plane components of spin, which
is independent of the $\Delta_{m}\sigma_{z}$ mass term.

\subsubsection*{Interference processes }

The carrier density coefficients vanish $\xi_{i\left(n\right)}^{bcd}\left(\omega\right)=0$.
The charge current coefficients are 
\begin{equation}
\begin{array}{rl}
\eta_{i\left(1\right)}^{abcd}\left(\omega\right)= & \frac{i\Theta\left(\omega-2\Delta_{m}\right)e^{4}A_{0}^{2}}{4\hbar^{3}\omega^{3}}\left(1-\frac{4\Delta_{m}^{2}}{\omega^{2}}\right)\left(\left[\frac{\left(\hat{\boldsymbol{a}}\cdot\hat{\boldsymbol{d}}\right)\hat{\boldsymbol{b}}\cdot\hat{\boldsymbol{c}}-2\left(\hat{\boldsymbol{a}}\cdot\hat{\boldsymbol{c}}\right)\hat{\boldsymbol{b}}\cdot\hat{\boldsymbol{d}}}{2}+\varphi^{abcd}\left(1-\frac{4\Delta_{m}^{2}}{\omega^{2}}\right)\right]+i\frac{2\Delta_{m}}{\omega}\left[\frac{\hat{\boldsymbol{a}}\cdot\hat{\boldsymbol{d}}\left(\hat{\boldsymbol{c}}\times\hat{\boldsymbol{b}}\right)\cdot\hat{\boldsymbol{z}}-2\hat{\boldsymbol{a}}\cdot\hat{\boldsymbol{c}}\left(\hat{\boldsymbol{d}}\times\hat{\boldsymbol{b}}\right)\cdot\hat{\boldsymbol{z}}}{2}\right]\right),\\
\eta_{i\left(2\right)}^{abcd}\left(\omega\right)= & \frac{i\Theta\left(\omega-\Delta_{m}\right)e^{4}A_{0}^{2}}{2\hbar^{3}\omega^{3}}\left(1-\frac{\Delta_{m}^{2}}{\omega^{2}}\right)\left(\left[\frac{\left(\hat{\boldsymbol{a}}\cdot\hat{\boldsymbol{c}}\right)\hat{\boldsymbol{b}}\cdot\hat{\boldsymbol{d}}+\left(\hat{\boldsymbol{a}}\cdot\hat{\boldsymbol{b}}\right)\hat{\boldsymbol{c}}\cdot\hat{\boldsymbol{d}}}{2}-2\varphi^{abcd}\left(1-\frac{\Delta_{m}^{2}}{\omega^{2}}\right)\right]+i\frac{\Delta_{m}}{\omega}\left[\frac{\hat{\boldsymbol{a}}\cdot\hat{\boldsymbol{c}}\left(\hat{\boldsymbol{d}}\times\hat{\boldsymbol{b}}\right)\cdot\hat{\boldsymbol{z}}+\hat{\boldsymbol{a}}\cdot\hat{\boldsymbol{b}}\left(\hat{\boldsymbol{d}}\times\hat{\boldsymbol{c}}\right)\cdot\hat{\boldsymbol{z}}}{2}\right]\right),
\end{array}
\end{equation}
\end{widetext} due to Eq. \eqref{eq:relations}, the spin density
coefficients can be written in terms of the ones for the charge current
as 
\begin{equation}
\zeta_{i\left(n\right)}^{abcd}\left(\omega\right)=\frac{\hbar}{2eA_{0}}\eta_{i\left(n\right)}^{\left(z\times a\right)bcd}\left(\omega\right),
\end{equation}
and the spin current coefficients can also be written in terms of
the ones for the charge current as 
\begin{equation}
\mu_{i\left(n\right)}^{abcde}\left(\omega\right)=\frac{\hbar D_{0}\Delta_{m}\hat{\boldsymbol{z}}\cdot\hat{\boldsymbol{a}}}{eA_{0}^{2}}\eta_{i\left(n\right)}^{bcde}\left(\omega\right),
\end{equation}
which finishes the list of optical injection coefficients.

\section{Optical injection coefficients for linear and circular polarizations }

\label{app:coeffs}

The coefficients used for one and two photons absorption processes
are 
\begin{equation}
\begin{array}{rl}
\xi_{1}^{xx}\left(\omega\right)= & \frac{\Theta\left(\omega-2\Delta_{m}\right)e^{2}}{8\hbar^{2}\omega}\left(1+\frac{4\Delta_{m}^{2}}{\omega^{2}}\right),\\
\xi_{2}^{xxxx}\left(\omega\right)= & \frac{\Theta\left(\omega-\Delta_{m}\right)e^{4}A_{0}^{2}}{4\hbar^{4}\omega^{5}}\left(1-\frac{\Delta_{m}^{2}}{\omega^{2}}\right)\left(1+\frac{3\Delta_{m}^{2}}{\omega^{2}}\right),
\end{array}
\end{equation}
and 
\begin{equation}
\begin{array}{rl}
\mu_{1}^{xyxx}\left(\omega\right)= & \frac{\Theta\left(\omega-2\Delta_{m}\right)e^{2}D_{0}}{8\hbar A_{0}}\left(1-\frac{4\Delta_{m}^{2}}{\omega^{2}}\right)\left(\frac{3}{4}+\frac{\Delta_{m}^{2}}{\omega^{2}}\right),\\
\mu_{1}^{yxxx}\left(\omega\right)= & -\frac{\Theta\left(\omega-2\Delta_{m}\right)e^{2}D_{0}}{8\hbar A_{0}}\left(1-\frac{4\Delta_{m}^{2}}{\omega^{2}}\right)\left(\frac{1}{4}+\frac{3\Delta_{m}^{2}}{\omega^{2}}\right),\\
\mu_{2}^{xyxxxx}\left(\omega\right)= & \frac{\Theta\left(\omega-\Delta_{m}\right)e^{4}D_{0}A_{0}}{4\hbar^{3}\omega^{4}}\left(1-\frac{\Delta_{m}^{2}}{\omega^{2}}\right)^{2}\left(1+\frac{\Delta_{m}^{2}}{\omega^{2}}\right),\\
\mu_{2}^{yxxxxx}\left(\omega\right)= & -\frac{\Theta\left(\omega-\Delta_{m}\right)e^{4}D_{0}A_{0}}{4\hbar^{3}\omega^{4}}\left(1-\frac{\Delta_{m}^{2}}{\omega^{2}}\right)^{2}\left(1+\frac{5\Delta_{m}^{2}}{\omega^{2}}\right),
\end{array}
\end{equation}
for linear polarization.

For circular polarization we have 
\begin{equation}
\begin{array}{rl}
\xi_{1}^{-\tau,+\tau}\left(\omega\right)= & \frac{\Theta\left(\omega-2\Delta_{m}\right)e^{2}}{8\hbar^{2}\omega}\left(1+\tau\frac{2\Delta_{m}}{\omega^{2}}\right)^{2},\\
\xi_{2}^{--++}\left(\omega\right)= & \frac{\Theta\left(\omega-\Delta_{m}\right)e^{4}A_{0}^{2}}{2\hbar^{4}\omega^{5}}\left(1-\frac{\Delta_{m}^{2}}{\omega^{2}}\right)\left(1+\tau\frac{\Delta_{m}}{\omega}\right)^{2},
\end{array}
\end{equation}
and 
\begin{equation}
\begin{array}{rl}
\mu_{1}^{-+-+}\left(\omega\right)= & \frac{i\Theta\left(\omega-2\Delta_{m}\right)e^{2}D_{0}}{16\hbar A_{0}}\left(1-\frac{4\Delta_{m}^{2}}{\omega^{2}}\right)\left(1+\tau\frac{2\Delta_{m}}{\omega}\right)^{2},\\
\mu_{2}^{-+--++}\left(\omega\right)= & \frac{i\Theta\left(\omega-\Delta_{m}\right)e^{4}D_{0}A_{0}}{2\hbar^{3}\omega^{4}}\left(1-\frac{\Delta_{m}^{2}}{\omega^{2}}\right)^{2}\left(1+\tau\frac{\Delta_{m}}{\omega}\right)^{2},
\end{array}
\end{equation}
also $\mu_{1}^{+--+}\left(\omega\right)=-\mu_{1}^{-+-+}\left(\omega\right)$
and $\mu_{2}^{-+--++}\left(\omega\right)=-\mu_{2}^{-+--++}\left(\omega\right)$.

The interference coefficients are 
\begin{equation}
\begin{array}{rl}
\eta_{i\left(1\right)}^{xxxx}\left(\omega\right)= & \frac{-i\Theta\left(\omega-2\Delta_{m}\right)e^{4}A_{0}^{2}}{32\hbar^{3}\omega^{3}}\left(1-\frac{4\Delta_{m}^{2}}{\omega^{2}}\right)\left(1+\frac{12\Delta_{m}^{2}}{\omega^{2}}\right),\\
\eta_{i\left(2\right)}^{xxxx}\left(\omega\right)= & \frac{i\Theta\left(\omega-\Delta_{m}\right)e^{4}A_{0}^{2}}{8\hbar^{3}\omega^{3}}\left(1-\frac{\Delta_{m}^{2}}{\omega^{2}}\right)\left(1+\frac{3\Delta_{m}^{2}}{\omega^{2}}\right),
\end{array}
\end{equation}
and 
\begin{equation}
\begin{array}{rl}
\eta_{i\left(1\right)}^{xxxy}\left(\omega\right)= & -\frac{\Theta\left(\omega-2\Delta_{m}\right)e^{4}A_{0}^{2}}{2\hbar^{3}\omega^{3}}\left(1-\frac{4\Delta_{m}^{2}}{\omega^{2}}\right)\frac{\Delta_{m}}{\omega},\\
\eta_{i\left(2\right)}^{xxxy}\left(\omega\right)= & \frac{\Theta\left(\omega-\Delta_{m}\right)e^{4}A_{0}^{2}}{2\hbar^{3}\omega^{3}}\left(1-\frac{\Delta_{m}^{2}}{\omega^{2}}\right)\frac{\Delta_{m}}{\omega},
\end{array}\label{eq:currcoeff1}
\end{equation}
also 
\begin{equation}
\begin{array}{rl}
\eta_{i\left(1\right)}^{yxxy}\left(\omega\right)= & \frac{i\Theta\left(\omega-2\Delta_{m}\right)e^{4}A_{0}^{2}}{32\hbar^{3}\omega^{3}}\left(1-\frac{4\Delta_{m}^{2}}{\omega^{2}}\right)\left(5-\frac{4\Delta_{m}^{2}}{\omega^{2}}\right),\\
\eta_{i\left(2\right)}^{yxxy}\left(\omega\right)= & \frac{-i\Theta\left(\omega-\Delta_{m}\right)e^{4}A_{0}^{2}}{8\hbar^{3}\omega^{3}}\left(1-\frac{\Delta_{m}^{2}}{\omega^{2}}\right)^{2},
\end{array}\label{eq:currcoeff2}
\end{equation}
and 
\begin{equation}
\begin{array}{rl}
\eta_{i\left(1\right)}^{+--+}\left(\omega\right)= & \frac{i\Theta\left(\omega-2\Delta_{m}\right)e^{4}A_{0}^{2}}{4\hbar^{3}\omega^{3}}\left(1-\frac{4\Delta_{m}^{2}}{\omega^{2}}\right)\left[\frac{1}{4}-\left(1+\tau\frac{\Delta_{m}}{\omega}\right)^{2}\right],\\
\eta_{i\left(2\right)}^{+--+}\left(\omega\right)= & \frac{i\Theta\left(\omega-\Delta_{m}\right)e^{4}A_{0}^{2}}{4\hbar^{3}\omega^{3}}\left(1-\frac{\Delta_{m}^{2}}{\omega^{2}}\right)\left(1+\tau\frac{\Delta_{m}}{\omega}\right)^{2},
\end{array}
\end{equation}
from which the other injection coefficients are obtained.



\begin{thebibliography}{10}
\bibitem{hasan10} M. Z. Hasan, and C. L. Kane, Rev. Mod. Phys. 82,
3045 (2010).

\bibitem{qi11} X.-L. Qi, and S.-C. Zhang, Rev. Mod. Phys. 83, 1057
(2011).

\bibitem{qi08} X.-L. Qi, T. L. Hughes, and S.-C. Zhang, Phys. Rev.
B 78, 195424 (2008).

\bibitem{essin09} A. M. Essin, J. E. Moore, and D. Vanderbilt, Phys.
Rev. Lett. 102, 146805 (2009).

\bibitem{fu08} L. Fu, and C. L. Kane, Phys. Rev. Lett. 100, 096407
(2008).

\bibitem{raghu10} S. Raghu, S. B. Chung, X.-L. Qi, and S.-C. Zhang,
Phys. Rev. Lett. 104, 116401 (2010).

\bibitem{butch12} C. Ojeda-Aristizabal, M. S. Fuhrer, N. P. Butch,
J. Paglione, and I. Appelbaum, Appl. Phys. Lett. 101, 023102 (2012).

\bibitem{modak12} S. Modak, K. Sengupta, and D. Sen, Phys. Rev. B
86, 205114 (2012).

\bibitem{semenov12} Y. G. Semenov, X. Li, and K. W. Kim, Phys. Rev.
B 86, 201401(R) (2012).

\bibitem{mahfouzi12} F. Mahfouzi, N. Nagaosa, and B. K. Nikoli\'{c},
Phys. Rev. Lett. 109, 166602 (2012).

\bibitem{tse10} W.-K. Tse and A. H. MacDonald, Phys. Rev. Lett. 105,
057401 (2010).

\bibitem{hosur11} P. Hosur, Phys. Rev. B 83, 035309 (2011).

\bibitem{misawa11} T. Misawa, T. Yokoyama, and S. Murakami, Phys.
Rev. B 84, 165407 (2011).

\bibitem{mciver12nn} J. W. McIver, D. Hsieh, H. Steinberg, P. Jarillo-Herrero,
and N. Gedik, Nature Nano. 7, 96 (2012).

\bibitem{refael13} A. Junck, G. Refael, and F. von Oppen, Phys. Rev.
B 88, 075144 (2013).

\bibitem{hsieh11a} D. Hsieh, J. W. McIver, D. H. Torchinsky, D. R.
Gardner, Y. S. Lee, and N. Gedik, Phys. Rev. Lett. 106, 057401 (2011).

\bibitem{hsieh11b} D. Hsieh, F. Mahmood, J.W. McIver, D. R. Gardner,
Y. S. Lee, and N. Gedik, Phys. Rev. Lett. 107, 077401 (2011).

\bibitem{mciver12prb} J. W. McIver, D. Hsieh, S. G. Drapcho, D. H.
Torchinsky, D. R. Gardner, Y. S. Lee, and N. Gedik, Phys. Rev. B 86,
035327 (2012).

\bibitem{sobota12} J. A. Sobota, S. Yang, J. G. Analytis, Y. L. Chen,
I. R. Fisher, P. S. Kirchmann, and Z.-X. Shen, Phys. Rev. Lett. 108,
117403 (2012).

\bibitem{rioux12} J. Rioux, and J. E. Sipe, Physica E 45, 1-15 (2012).

\bibitem{kiran11} K. M. Rao, and J. E. Sipe, Phys. Rev. B 84, 205313
(2011).

\bibitem{norris10} D. Sun, C. Divin, J. Rioux, J. E. Sipe, C. Berger,
W. A. de Heer, P. N. First, and T. B. Norris, Nano Lett. 10, 1293-1296
(2010).

\bibitem{rioux11} J. Rioux, G. Burkard, and J. E. Sipe, Phys. Rev.
B 83, 195406 (2011).

\bibitem{kiran12} K. M. Rao, and J. E. Sipe, Phys. Rev. B 86, 115427
(2012).

\bibitem{virk11} K. S. Virk, and J. E. Sipe, Phys. Rev. Lett. 107,
120403 (2011).

\bibitem{liu10} C.-X. Liu, X.-L. Qi, H. Zhang, X. Dai, Z. Fang, and
S.-C. Zhang, Phys. Rev. B 82, 045122 (2010).

\bibitem{lu10} H.-Z. Lu, W.-Y. Shan, W. Yao, Q. Niu, and S.-Q. Shen,
Phys. Rev. B 81, 115407 (2010).

\bibitem{driel06prl} H. Zhao, E. J. Loren, H. M. van Driel, and A.
L. Smirl, Phys. Rev. Lett. 96, 246601 (2006).

\bibitem{driel06ssc} E. Ya. Sherman, A. Najmaie, H. M. van Driel,
A. L. Smirl, J.E. Sipe, Solid State Comm. 139, 439-446 (2006).\end{thebibliography}
\end{document}